%% file: main.tex
\documentclass[manuscript, nonacm]{acmart}
\usepackage{xcolor}
\usepackage{subcaption}
\usepackage[skins]{tcolorbox}
\tcbuselibrary{skins} 
\usepackage{tabularx}
\usepackage{tcolorbox}
\usepackage{algorithm}
\usepackage{booktabs}
\usepackage{booktabs,tablefootnote, threeparttable, colortbl, multirow}
\definecolor{mycolorsummary}{HTML}{ACD6EC}
\usepackage{xspace}
\newcommand{\DeepGroundRepair}{\mbox{\textsc{DeepGroundRepair}}\xspace}
\usepackage{pifont}

\AtBeginDocument{%
  }

\setcopyright{none}
\begin{document}

\title{\DeepGroundRepair: From API-Grounded Generation to Shape-Aware Repair for Deep Learning Programs}

\author{Xiangyue Ma}
\affiliation{%
  \institution{Beihang University}
  \city{Beijing}
  \country{China}}
\email{xiangy\_ma@buaa.edu.cn}
\orcid{0009-0008-1155-905X}

\author{Zhen Wang}
\affiliation{%
  \institution{Beihang University}
  \city{Beijing}
  \country{China}}
\email{wz2603520@buaa.edu.cn}

\author{Xuhui Lu}
\affiliation{%
  \institution{Beihang University}
  \city{Beijing}
  \country{China}}
\email{luxuhui@buaa.edu.cn}
\orcid{0000-0001-6839-5633}

\author{Xiaoting Du}
\affiliation{%
  \institution{Beijing University of Posts and Telecommunications}
  \city{Beijing}
  \country{China}}
\email{duxiaoting@bupt.edu.cn}
\orcid{0000-0002-1609-0480}

\author{Chenglong Li}
\affiliation{%
  \institution{Beijing University of Technology}
  \city{Beijing}
  \country{China}}
\email{lichenglong@bjut.edu.cn}
\orcid{0000-0001-5777-6828}

\author{Jun Li}
\affiliation{%
  \institution{Beihang University}
  \city{Beijing}
  \country{China}}
\email{2503515@buaa.edu.cn}

\author{Haotian Li}
\affiliation{%
  \institution{Beihang University}
  \city{Beijing}
  \country{China}}
\email{lihaotian08@buaa.edu.cn}

\author{Zheng Zheng}
\authornote{Corresponding author}
\affiliation{%
  \institution{Beihang University}
  \city{Beijing}
  \country{China}}
\email{zhengz@buaa.edu.cn}
\orcid{0000-0001-7922-9067}

\renewcommand{\shortauthors}{Ma et al.}

\begin{abstract}
  Deep learning (DL) programs are fundamental to modern AI applications, and automatically generating them from natural-language requirements can substantially reduce DL development effort. Large language models (LLMs) have demonstrated promising capabilities for this direction. However, generated DL programs often remain unreliable because they must satisfy domain-specific API usage constraints and tensor-shape dependencies, violations of which can cause API-level syntactic issues and runtime failures. Existing retrieval-augmented generation and feedback-guided code repair methods do not adequately address these DL-specific constraints.

  In this paper, we propose \DeepGroundRepair, a three-stage framework that establishes a generation--localization--repair workflow to improve the reliability of LLM-generated DL programs. Its key idea is to progressively narrow the generation and repair spaces using DL-specific knowledge and runtime tensor information. Specifically, it grounds generation in task-relevant API knowledge, localizes the upstream causes of tensor-shape mismatch failures, and uses runtime tensor shapes to guide parameter repair. Together, these mechanisms transform open-ended generation and repair into a knowledge-grounded and shape-aware process.

  We evaluate \DeepGroundRepair on the 100 model-level DL programming tasks in DeepEval using six open-weight LLMs across TensorFlow, PyTorch, and PaddlePaddle. We compare it with four baselines: two generation-only methods and their counterparts incorporating traceback-guided self-debugging. The results show that, relative to these baselines, \DeepGroundRepair reduces the combined number of API-level syntactic issues by an average of 5.4 per 10 program files and achieves an average relative improvement of 85.3\% in code executability. Compared with the two self-debugging baselines, it improves semantic consistency by an average of 18.1\% and reduces token cost by an average of 28.4\%. The consistent improvements across model families, model scales, and DL libraries provide evidence of its generalizability within the evaluated scope.
\end{abstract}




\maketitle

\section{Introduction} \label{sec:introduction}
In recent years, large language models (LLMs) have demonstrated strong capabilities in generating code from natural language~\cite{Survey,structuredcot,ROCODE}. By translating programming intent into executable code, LLMs can lower barriers to and reduce the effort required for software development. Yet, performance on general-purpose code generation does not necessarily extend to domain-specific programming tasks. This challenge is especially pronounced in DL programming, where structural and semantic dependencies make DL programs prone to bugs involving API usage and tensor compatibility~\cite{Taxonomyrealfaultsdeeplearningsystems,gDefects4DL,MisuAPI}. Therefore, improving the reliability of LLM-generated DL programs remains a challenge.

In DL programs, DL models are constructed as computation graphs of interdependent API calls and tensor transformations. Program correctness therefore depends on library-specific API configurations and tensor compatibility across the graph. As demonstrated in our prior study~\cite{DeepEval}, LLM-generated DL code frequently encounters two prevalent problems. \textbf{\textit{Problem 1}: API-Level Syntactic Issues.} LLMs frequently hallucinate or misuse DL APIs, including references to nonexistent members and violations of API signatures. \textbf{\textit{Problem 2}: Runtime Execution Failures.} LLM-generated DL code frequently fails at runtime due to violations of implicit constraints among DL operations. Among such failures, tensor‑shape mismatches are the predominant type, accounting for 77.34\% of execution failures on average~\cite{DeepEval}.

Existing studies relevant to these problems mainly follow two directions. One line of work employs retrieval-augmented generation (RAG) to ground code generation in external knowledge~\cite{DocPromptinggeneratingcoderetrievingdocs,WhenLLMsmeetAPIdocumentation,mitigatingcodeLLMhallucinationsAPIdocumentation}. Its effectiveness, however, depends heavily on retrieving task-relevant context, as irrelevant information can introduce noise and degrade code generation~\cite{Whatretrieveeffectiveretrieval}. In DL programming, this challenge is further complicated by the gap between high-level task intent and fine-grained API requirements, making static query--document matching prone to retrieving irrelevant API knowledge~\cite{su-etal-2024-evor}. Another line of work investigates feedback-guided code repair, in which failure information from program execution is fed back to LLMs for iterative refinement~\cite{codecorrectness,Selfedit,Teachinglargelanguagemodelsselfdebug}. However, prior evidence shows that such repair often yields only limited improvements~\cite{olausson2024self}. This limitation is particularly evident in DL programs, where raw tracebacks often reveal only downstream symptoms of tensor-shape errors. Consequently, an LLM reasoning solely from such feedback lacks the evidence needed to localize upstream root causes. Moreover, even after the root cause is localized, self-repair without explicit shape guidance remains insufficient: resolving such errors requires accounting for how multiple interdependent API parameters jointly affect tensor shapes~\cite{Deng2024CanLLMs}.

As discussed above, existing approaches have shown promise in improving general-purpose LLM-generated code, but they do not fully address the two DL-specific problems. Therefore, this work aims to improve the \textbf{syntactic correctness} (\textit{\textbf{Problem 1}}) and \textbf{runtime executability} (\textit{\textbf{Problem 2}}) of LLM-generated DL programs. To this end, we identify one key retrieval challenge for mitigating API-level syntactic issues during code \textbf{generation} and two key challenges concerning the \textbf{localization} and \textbf{repair} of tensor-shape mismatches underlying runtime execution failures.

\begin{itemize}
  \item \textbf{\textit{Challenge 1 (Generation)}: How to provide LLMs with task-relevant DL API knowledge?} High-level DL task requirements typically describe model structure and functionality rather than explicit API usage, creating a semantic gap between task intent and API documentation.
  \item \textbf{\textit{Challenge 2 (Localization)}: How to localize the root-cause API of a tensor-shape mismatch?} A tensor-shape mismatch may originate from an upstream API configuration but surface only at a downstream operation, separating the failure site from its root cause.
  \item \textbf{\textit{Challenge 3 (Repair)}: How to infer shape-consistent API parameters for repair?} Without explicit guidance on how interdependent API parameters jointly affect tensor shapes, LLMs struggle to identify configurations satisfying tensor-shape constraints.
\end{itemize}


To address these challenges, we propose \DeepGroundRepair, a three-stage framework that establishes a \textbf{generation-localization-repair} workflow for DL programs, progressing from API-grounded retrieval-augmented generation through root-cause localization to shape-aware repair. Its main idea is to progressively narrow the generation and repair spaces using DL API knowledge and runtime tensor shapes as complementary guidance. API knowledge grounds code generation, while runtime tensor shapes support root-cause localization and constrain subsequent parameter-level repair. The three stages work in concert to address the challenges outlined above.
(1) \textbf{\textit{Reasoning-Augmented Classification Retrieval}.} To accurately obtain task-relevant API knowledge, \DeepGroundRepair uses LLM-based API recommendation and existence verification to distinguish existing APIs from hallucinated ones, routing existing and nonexistent API candidates to sparse and dense retrieval, respectively.
(2) \textbf{\textit{Root-Cause API Localization}.} To localize where a tensor shape mismatch should be repaired, \DeepGroundRepair combines static program analysis with dynamic execution traces to capture both program structure and runtime behavior. This combined analysis traces the mismatch from its failure site back to the upstream root-cause API and masks suspicious parameters.
(3) \textbf{\textit{Shape-Aware Self-Repair}.} To determine how the localized API should be repaired, \DeepGroundRepair uses traced tensor shapes as explicit evidence and provides the LLM with retrieved shape-related knowledge for the root-cause API. This guidance enables the LLM to infer shape-consistent parameter values while respecting tensor-shape dependencies across the computation graph.

To evaluate the effectiveness of \DeepGroundRepair, we conduct experiments on the model-level DL code generation benchmark DeepEval using six open-weight LLMs from two model families across TensorFlow, PyTorch, and PaddlePaddle. We compare it against four baselines: two generation-only methods (vanilla LLM generation and Hybrid-RAG) and two methods combining their respective generation strategies with traceback-guided self-debugging. Across the three libraries, \DeepGroundRepair reduces API hallucinations and parameter misuses by an average of 2.14 and 3.26 issues per 10 program files, respectively. Most notably, it improves code executability by an average of 85.3\% relative to the four baselines. It also achieves an average relative improvement of 18.1\% in semantic consistency over the two self-debugging baselines. Compared with these baselines, \DeepGroundRepair further delivers higher code executability with 28.4\% lower token cost on average.

In summary, this work makes the following contributions:
\vspace{-0.54em}
\begin{itemize}
\item We propose \DeepGroundRepair, a novel framework that bridges API-grounded code generation and shape-aware repair, enabling DL programs to be progressively improved through task-relevant API knowledge and runtime tensor-shape feedback.
\item We conduct extensive experiments across multiple LLMs and DL libraries. The results demonstrate that \DeepGroundRepair consistently improves the syntactic correctness, code executability, and semantic consistency of LLM-generated DL programs, while achieving a favorable effectiveness--efficiency trade-off.
\item We perform ablation study and stage-specific analyses to quantify the contributions of the three stages and examine the mechanisms underlying the overall performance gains.
\item We release the replication package of \DeepGroundRepair to facilitate reproducibility and support future research on reliable LLM-based DL code generation.
\end{itemize}

\section{Motivating Examples}
\label{sec:motivating_example}
Despite advances in LLM-based code generation, producing reliable DL programs remains challenging because it requires both library-specific API knowledge and effective repair of tensor-shape mismatches. This section presents representative failures of LLM-generated DL programs and highlights the key challenges motivating \DeepGroundRepair.

\subsection{Problem 1: API-Level Syntactic Issues.}
LLM-generated DL programs frequently contain API-level syntactic issues, including API hallucination and API parameter misuse~\cite{DeepEval,DemystifyingDetectingMisusesDeepLearningAPIs}. Figure~\ref{fig:syntactic_issues_label} presents two representative examples. The first example demonstrates API hallucination, where the generated code invokes the nonexistent \texttt{torch.nn.Concatenate} API. The second example illustrates API parameter misuse, where \texttt{paddle.nn.Conv2D} is configured with the unsupported \texttt{channels} parameter while omitting the required \texttt{in\_channels} and \texttt{out\_channels} parameters.
\input{figures/motivation/code_syntax/syntactic_issues.tex}

\input{figures/motivation/code_syntax/retrieval_failure.tex}
These failures show that reliable DL code generation requires both identifying APIs that implement the intended operations and using them in accordance with their signatures and parameter specifications. Existing retrieval methods, however, often fail to provide such knowledge because task requirements express high-level modeling intents, whereas API documentation provides fine-grained, library-specific details. As shown in Figure~\ref{fig:retrieval_failure}, directly using the complete task requirement as a dense-retrieval query fails to bridge this gap. Despite the target architecture implicit in the requirement, most retrieved APIs are irrelevant, and only one of the six required APIs, \texttt{tensorflow.keras.layers.Softmax}, appears among the top 10 candidates. This limitation motivates Stage~1, \textit{Reasoning-Augmented Classification Retrieval}, which addresses our first challenge: \textbf{\textit{How to provide LLMs with task-relevant DL API knowledge?}}

\subsection{Problem 2: Runtime Execution Failures}
Beyond API-level syntactic issues, LLM-generated DL programs frequently encounter runtime failures. Our prior analysis of 3,677 failures across eight LLMs found tensor-shape mismatches to be the predominant failures type, accounting for 77.34\% of execution failures on average~\cite{DeepEval}. Such mismatches arise from violations of tensor-shape constraints during execution. As illustrated in Figure~\ref{fig:executability_failure_label}, the generated program fails when the flattened feature tensor is passed to the fully connected layer. Specifically, the flattened tensor contains 401,408 features, whereas the configured \texttt{in\_features} of \texttt{nn.Linear} is 25,088, leading to a shape incompatibility at the layer interface.
\input{figures/motivation/code_executability/executability_failure.tex}

This failure highlights two key challenges in tensor-shape-aware repair. First, the location where a mismatch is detected does not necessarily correspond to the API configuration that introduces the incompatible shape. In this example, the execution failure is reported at \texttt{x=self.fc1(x)}, whereas the faulty parameter configuration is defined earlier when constructing the \texttt{nn.Linear} layer in \texttt{\_\_init\_\_}. Therefore, execution feedback alone may be insufficient for root-cause localization, whereas tracing tensor propagation can help identify the API configuration responsible for the mismatch. This observation motivates the design of Stage~2, \textit{Root-Cause Localization}, to address the challenge: \textbf{\textit{How to localize the root-cause API of a tensor-shape mismatch?}}

Second, after localizing the root-cause API, repairing a mismatch requires inferring parameter values that satisfy tensor-shape constraints rather than performing unconstrained parameter repair. However, runtime tracebacks offer little guidance on how API parameters govern tensor transformations. Moreover, prior studies have shown that LLMs alone struggle to reason about complex numerical constraints among API parameters~\cite{Deng2024CanLLMs}. Together, these limitations highlight the need for explicit shape guidance in valid parameter inference and motivate the design of Stage~3, \textit{Shape-Aware Self-Repair}, to address the challenge: \textbf{\textit{How to infer shape-consistent API parameters for repair?}}

The above examples reveal that generation and repair of LLM-generated DL programs requires bridging the task--API knowledge gap, localizing root-cause API configurations, and inferring shape-consistent parameters. Together, these examples also motivate the generation--localization--repair workflow of \textsc{DeepGroundRepair}, which is introduced in the next section.

\input{figures/overview.tex}
\section{Proposed Approach} \label{sec: proposed approach}
To address the challenges identified above, we propose \DeepGroundRepair, a three-stage framework that implements a generation--localization--repair workflow for DL programs. As illustrated in Figure~\ref{fig:overview}, the framework progresses from \textit{Reasoning-Augmented Classification Retrieval} (Stage~1) for API-grounded code generation, through \textit{Root-Cause Localization} (Stage~2) for narrowing the repair space, to \textit{Shape-Aware Self-Repair} (Stage~3) for deriving shape-consistent repairs. Together, these stages are designed to reduce API-level syntactic issues and improve code executability while preserving semantic consistency.

To support Stages~1 and~3, \DeepGroundRepair constructs a library-specific API documentation database shared across the two stages. Rather than providing complete API documentation, the framework selectively retrieves knowledge tailored to each stage: Stage~1 uses API names, signatures, and descriptions to ground draft code generation, whereas Stage~3 uses the parameter description and tensor-shape transformation formulas of the localized root-cause API to guide repair. Details of database construction are provided in Section~\ref{sec:implementation}.

The three stages form an integrated pipeline, with the output of each preceding stage facilitating the next. The core idea and design objective of each stage are summarized below.

\textbf{(1) Stage~1: Reasoning-Augmented Classification Retrieval.}
This stage aims to reduce factual hallucinations in DL API usage by grounding code generation in task-relevant API knowledge. \DeepGroundRepair first recommends the APIs required by the task and verifies their existence. Verified APIs are routed to sparse retrieval, whereas suspected hallucinations are routed to dense retrieval. The retrieved knowledge is then provided to the LLM to generate a more API-compliant draft program. Programs with tensor-shape mismatches proceed to Stage~2, whereas the few other execution failures are handled using the single-round traceback-guided strategy adopted by the self-debugging baselines.

\textbf{(2) Stage~2: Root Cause Localization.}
Given the generated program, this stage narrows the repair space by identifying the root-cause API and masking its suspicious parameters. \DeepGroundRepair combines abstract syntax tree (AST)-based static analysis with dynamic execution tracing to capture program structure and runtime tensor-shape propagation. The resulting parameter-masked program and traced tensor shapes provide focused context for shape-aware repair in Stage~3.

\textbf{(3) Stage~3:  Shape-Aware Self-Repair.}
Based on the localization results from Stage~2, this stage guides the LLM to self-repair the buggy program using explicit shape evidence and knowledge. The program is annotated with the traced shapes, which serve as explicit evidence for the LLM. Shape-related knowledge of the root-cause API, including parameter descriptions and tensor-shape transformation formulas, is then retrieved to augment the repair context. Combining this knowledge with runtime shape evidence, the LLM infers shape-consistent parameter values and uses them to resolve the tensor-shape mismatch.

\input{algorithm/stage1.tex}
\subsection{Stage~1: Reasoning-Augmented Classification Retrieval} \label{sec:stage1}
Naive RAG directly retrieves API documentation using high-level task requirements, leaving a semantic gap between DL programming intent and fine-grained API knowledge. Moreover, static query--document matching cannot reliably uncover the APIs implicitly required to implement complex DL model architectures. To bridge this gap, this stage prompts the LLM to reason over the task requirement and identify candidate APIs implicitly required to implement the target DL models. It then verifies these candidates and routes existing and hallucinated candidates to sparse and dense retrieval, respectively.
Algorithm~\ref{alg:stage1} summarizes the workflow of this stage, and the individual steps are detailed below.

\textbf{(1) LLM-Based Reasoning for API Recommendation.} 
To address the semantic gap between task requirements and API documentation, \DeepGroundRepair prompts the LLM to reason about which library-specific APIs are implicitly required by each task. The LLM then produces a structured list of candidate API names, translating the task description into an intermediate representation closer to the API documentation space. For reproducibility, we provide the prompt template used for API recommendation in Figure~\ref{fig:prompt1}.

\textbf{(2) API Existence Verification.} 
After obtaining the API names recommended by the LLM, \DeepGroundRepair verifies each candidate against the target DL library. The probabilistic nature of LLM generation introduces uncertainty into recommendations. Moreover, LLM parametric knowledge of library-specific APIs may be imprecise, particularly for less widely used libraries such as PaddlePaddle. Consequently, the recommendations may include hallucinated API names. Directly using such candidates as queries may return task-irrelevant API documentation, thereby misguiding the LLM during subsequent code generation. To distinguish hallucinated candidates from existing APIs before retrieval, \DeepGroundRepair resolves module paths and members using \texttt{importlib.import\_module} and \texttt{getattr}, respectively.

\textbf{(3) Classification Retrieval.}
Based on API existence verification, \DeepGroundRepair routes each candidate API to either sparse or dense retrieval~\cite{luan2021sparse}. This strategy avoids misleading lexical matches caused by hallucinated API names while preserving their semantic information as retrieval clues.

For verified APIs, their canonical names enable exact lexical matching. Sparse retrieval ranks documents based on lexical overlap with a query. Accordingly, \DeepGroundRepair uses sparse retrieval to retrieve the corresponding API entries from the knowledge base. Given a candidate API name $q$ and an API entry $d$, their sparse relevance score is computed as:

\begin{equation}
\operatorname{score}_{\mathrm{sparse}}(q,d)
=
\sum_{t \in q}
\operatorname{IDF}(t)
\cdot
\frac{f(t,d)(k_1+1)}
{f(t,d)+k_1\left(1-b+b\frac{|d|}{\mathrm{avgdl}}\right)},
\label{eq:bm25}
\end{equation}
where $f(t,d)$ denotes the frequency of term $t$ in $d$, and $k_1$ and $b$ are BM25 parameters~\cite{robertson2009probabilistic}.

For candidate API names determined to be non-existent, lexical differences from valid API names may obscure relevant matches. Dense retrieval ranks documents based on their proximity to the query in an embedding space. Accordingly, \DeepGroundRepair uses dense retrieval to retrieve semantically similar valid APIs. The dense relevance score between $q$ and $d$ is computed as follows:
\begin{equation}
\operatorname{score}_{\mathrm{dense}}(q,d)
=
-\left\|E(q)-E(d)\right\|_2^2,
\label{eq:dense}
\end{equation}
where $E(\cdot)$ denotes the embedding model. Higher scores indicate greater semantic similarity between the candidate API and the retrieved API entry.

\textbf{(4) LLM-based Draft Code Generation.}
After classification retrieval, \DeepGroundRepair obtains a set of structured API documents. Rather than providing the full documents to the LLM, Stage~1 retains only the API names, signatures, and descriptions as grounding context for code generation. These fields capture the APIs' core functionality and parameter specifications. The resulting shorter prompt also helps the LLM focus on essential information~\cite{liu2024lost}. Guided by the retrieved API knowledge, the LLM generate a draft DL program following the prompt template shown in Figure~\ref{fig:prompt2}.
 
\subsection{Stage~2: Root-Cause Localization}
\input{algorithm/stage2.tex}
Despite API-grounded generation, the draft program may still contain tensor-shape faults that emerge only at runtime. These faults are difficult to diagnose because layer parameters defined in the model initialization method (\texttt{\_\_init\_\_}) govern tensor transformations in the forward-computation method (\texttt{forward} or \texttt{call})\footnote{\texttt{forward} is used in PyTorch and PaddlePaddle, whereas \texttt{call} is used in TensorFlow.}, while the resulting mismatches may surface only at downstream operations. Consequently, a traceback typically identifies only the crash site, not the upstream API and parameters causing the mismatch. To address this limitation, Stage~2 backtracks tensor propagation from the crash line to localize the root-cause API and mask its suspicious parameters. This process narrows the repair space and provides focused guidance for Stage~3. Algorithm~\ref{alg:stage2} summarizes the workflow of this stage, and the individual steps are detailed below.

\textbf{Tensor Shape Tracing.}
This step reconstructs runtime tensor propagation by tracing intermediate tensor shapes throughout forward execution. \DeepGroundRepair first performs AST-based static analysis of the generated program to identify function calls and tensor variables involved in forward computation as candidate observation points. It then executes the program with line-level tracing using \texttt{sys.settrace}, recording tensor shapes at these points until an exception occurs. The resulting trace provides explicit tensor-shape context for Stage~3 and auxiliary evidence for root-cause localization.

\textbf{Root-Cause API Localization.}
When a crash occurs, \DeepGroundRepair locates the failing operation from the traceback and uses AST-based data-flow analysis to trace the involved tensors backward through the forward computation. The recorded shapes reveal how tensor dimensions change along this dependency path, helping identify the layer configuration inconsistent with runtime tensor propagation. The framework then maps the relevant layer instance to its constructor call in \texttt{\_\_init\_\_} and identifies the corresponding API as the root cause. The Stage~2 panel of Figure~\ref{fig:overview} illustrates this process using an excerpt from the tensor-mismatch example in Figure~\ref{fig:executability_failure_label}. Specifically, the traceback points to \texttt{self.fc1(x)} in \texttt{forward}, while the shape trace records the dimensions of \texttt{x}. The AST analysis then maps \texttt{self.fc1} to the \texttt{nn.Linear} constructor assigned to it in \texttt{\_\_init\_\_}.

\textbf{Buggy Parameter Masking.}
Once the root-cause API is localized, \DeepGroundRepair selects and masks its potentially responsible parameters. It first maps library-specific layer APIs to unified semantic layer categories, such as convolutional, pooling, and fully connected layers. Each semantic layer category has predefined masking rules for channel and spatial mismatches, specifying which API parameters control the corresponding tensor dimensions. \DeepGroundRepair then infers the mismatch type from the traced tensor shapes and applies the corresponding rule for the root-cause API's category. For a channel mismatch, the rule masks parameters governing the expected number of input or output channels, as illustrated by the masked \texttt{in\_features} argument in the Stage~2 panel of Figure~\ref{fig:overview}. For a spatial mismatch, it masks parameters governing the output height and width, as illustrated by the masked \texttt{kernel\_size}, \texttt{strides}, and \texttt{padding} arguments of the root-cause \texttt{self.conv\_branch} layer in Figure~\ref{fig:inter_path_mismatch}.

\input{algorithm/stage3.tex}
\subsection{Stage~3: Shape-Aware Self-Repair} \label{sec:stage3}
Building on the root-cause API and runtime tensor shapes obtained in Stage~2, this stage infers shape-consistent values for the masked parameters. The relevant shape constraints span different dependency scopes. Accordingly, \DeepGroundRepair distinguishes between \textit{intra-path mismatches} and \textit{inter-path mismatches} and applies mismatch-specific repair strategies. Intra-path mismatches arise between operations along a single computation path, whereas inter-path mismatches arise when tensors from different paths cannot be merged because of incompatible shapes. Algorithm~\ref{alg:stage3} summarizes the workflow, whose individual steps are described below.

\textbf{Tensor Shape Annotation.} 
\DeepGroundRepair annotates relevant statements in the forward computation with the corresponding tensor shapes traced in Stage~2. In the Stage~3 panel of Figure~\ref{fig:overview}, \texttt{x=self.flatten(x)} is annotated with its output shape \texttt{(1,3072)}, making the local input-shape constraint of the subsequent \texttt{self.fc1} call explicit. As illustrated in Figure~\ref{fig:inter_path_mismatch}, \texttt{branch=self.conv\_branch(inputs)} is annotated with its output shape \texttt{(1,28,28,16)} and the target shape \texttt{(1,14,14,16)} required for merging with the main-path tensor at \texttt{Add}. The resulting annotations expose the local or cross-path shape constraints to be satisfied during repair.

\textbf{Shape-Related API Knowledge Retrieval.} 
For intra-path mismatches, the annotated local shape context is generally sufficient for repair. For inter-path mismatches, however, the annotations specify the required cross-path transformation but not the parameter values that produce it. For example, Figure~\ref{fig:inter_path_mismatch} shows that the masked \texttt{kernel\_size}, \texttt{strides}, and \texttt{padding} of \texttt{conv\_branch} must be inferred so that its output matches the main-path tensor before \texttt{Add}.
\input{figures/motivation/code_executability/inter_path_mismatch.tex}

To relate API parameters to the target tensor shape, \DeepGroundRepair retrieves parameter descriptions and, when available, tensor-shape transformation formulas for the root-cause API. As shown in Figure~\ref{fig:shape_related_knowledge}, the retrieved knowledge for \texttt{tensorflow.keras.layers.Conv2D} comprises a structured parameter--description mapping and the corresponding transformation formulas. The semantic mapping indicates that \texttt{kernel\_size}, \texttt{strides}, and \texttt{padding} govern the output spatial dimensions, whereas \texttt{filters} determines the output channels. Under the channels-last format\footnote{$N$ denotes batch size, while $H$, $W$, and $C$ denote height, width, and channel count, respectively. Subscripts $\mathrm{in}$ and $\mathrm{out}$ distinguish input and output.}, the formulas characterize how the tensor shape transforms from $(N,H_{\mathrm{in}},W_{\mathrm{in}},C_{\mathrm{in}})$ to $(N,H_{\mathrm{out}},W_{\mathrm{out}},C_{\mathrm{out}})$ as governed by the corresponding API parameters. Together, the parameter descriptions and tensor shape transformation formulas guide the LLM in coordinating parameter values to satisfy cross-path shape constraints.

\input{tables/method/shape_related_knowledge.tex}

\textbf{Shape-Aware Code Repair.}
To account for their different dependency scopes, \DeepGroundRepair adopts distinct repair strategies for intra-path and inter-path mismatches. For intra-path mismatches, \DeepGroundRepair prompts the LLM to generate a repaired code line from the shape-annotated and parameter-masked program. The generated line then replaces the corresponding masked API call in the original code. 

For inter-path mismatches, \DeepGroundRepair adopts a generate-and-validate strategy. In the example shown in Figure~\ref{fig:inter_path_mismatch}, the shape of the main-path tensor entering \texttt{Add} defines the target shape for the output of the shortcut-path \texttt{Conv2D}. Using annotations that capture this target shape and tensor-shape propagation process, together with retrieved shape-related knowledge for \texttt{Conv2D}, the LLM proposes candidate combinations for the masked parameter values. A lightweight constraint solver then evaluates each combination by checking whether the resulting \texttt{Conv2D} output matches the target shape. Finally, the validated combination, comprising \texttt{kernel\_size=(3,3)}, \texttt{strides=(2,2)}, and \texttt{padding="same"}, is inserted into the masked \texttt{Conv2D} call to produce the repaired code. For reproducibility, the prompt templates for intra-path and inter-path repair are provided in Figures~\ref{fig:prompt3} and~\ref{fig:prompt4}, respectively.
 
\section{Experimental Setup}
To systematically evaluate the proposed \DeepGroundRepair framework, we organize our experimental study around the following five research questions (RQs). We also describe the baselines, evaluation metrics, benchmark, and implementation details used in our evaluation.

\begin{itemize}
  \item \textbf{RQ1: End-to-End Effectiveness.} How does \DeepGroundRepair perform end to end compared with the baselines?
  \item \textbf{RQ2: Ablation Study.} How much does each key component contribute to the end-to-end performance of \DeepGroundRepair?
  \item \textbf{RQ3: Stage-specific Analysis.} How effectively do the three stages of \DeepGroundRepair address the corresponding challenges in generation, localization, and repair for DL programs?
  \item \textbf{RQ4: Generalizability.} How well do the performance gains of \DeepGroundRepair generalize across different model families, model scales, and DL libraries?
  \item \textbf{RQ5: Token Efficiency.} How does \DeepGroundRepair balance effectiveness and token cost compared with the baselines?
\end{itemize}

In brief, we evaluate the \textbf{effectiveness} of \DeepGroundRepair in RQ1, examine component \textbf{necessity} through an ablation study in RQ2, analyze the underlying \textbf{mechanisms} via stage-specific analysis in RQ3, investigate the \textbf{generalizability} of our approach in RQ4, and assess its \textbf{efficiency} in RQ5. For evaluation, we employ end-to-end effectiveness metrics for RQ1, RQ2, and RQ4, while stage-specific effectiveness metrics are used for RQ3. The detailed definitions of all metrics are provided in Section~\ref{sec:evaluation_metrics}.

\subsection{Baselines}
To evaluate the effectiveness of \DeepGroundRepair, we compare it with four baselines: two generation-only baselines (\textit{Baselines~1 and~2}) and their self-debugging counterparts (\textit{Baselines~3 and~4}), which perform one traceback-guided repair iteration upon execution failure.

\subsubsection{Baseline 1: Vanilla LLM}
This baseline uses zero-shot prompting to generate DL programs directly from the task requirements. It evaluates the model's fundamental ability to generate DL programs without retrieval augmentation or iterative repair.

\subsubsection{Baseline 2: Hybrid-RAG}
Following prior work~\cite{VulRAGenhancingLLMbasedvulnerabilitydetectionknowledgelevelRAG}, we adapt Hybrid-RAG as a baseline that combines sparse and dense retrievers to capture both lexical and semantic similarity. The two retrievers independently produce ranked lists of API documents. We fuse these lists using reciprocal rank fusion (RRF)~\cite{Reciprocalrankfusion}, which assigns each API document $d$ appearing in either list the following score:
\begin{equation}
\label{eq:rrf}
\mathrm{RRF}(d) =
\sum_{j=1}^{m}
\frac{w_j}{c + r_j(d)},
\end{equation}
where $w_j$ is the weight of the $j$-th retriever, $r_j(d)$ is the rank of document $d$ in its result list, and $c=60$ is the rank constant. Following standard RRF, we assign equal weights to both retrievers ($w_j=1$) and provide the top-10 fused documents to the LLM for code generation.

\subsubsection{Baseline 3: Vanilla LLM with Self-Debugging}
Prior work has shown that execution feedback can help LLMs diagnose and repair errors in generated programs~\cite{codecorrectness,Teachinglargelanguagemodelsselfdebug}. Following this feedback-guided paradigm, \textit{Baseline~3} augments vanilla LLM generation with traceback-guided self-debugging. It generates an initial code solution using the same setting as \textit{Baseline~1} and executes it. If execution fails, the LLM performs a repair iteration using the resulting runtime traceback as feedback.

\subsubsection{Baseline 4: Hybrid-RAG with Self-Debugging}
To evaluate the combined effectiveness of retrieval-augmented generation and feedback-guided repair, \textit{Baseline~4} integrates Hybrid-RAG-based code generation with traceback-guided self-debugging. It first uses Hybrid-RAG to generate an initial code solution and executes it. If execution fails, the LLM performs a repair iteration using the resulting runtime traceback as feedback.

\subsection{Evaluation Metrics} \label{sec:evaluation_metrics}
This section presents the evaluation metrics used to compare \DeepGroundRepair with baseline methods. The metrics are categorized into three groups according to their evaluation objectives: end-to-end effectiveness metrics for RQ1, RQ2, and RQ4; stage-specific effectiveness metrics for RQ3; and the token efficiency metric for RQ5.

\subsubsection{End-to-End Effectiveness Metrics} \label{sec:end_to_end_effectiveness_metrics}
To evaluate the overall effectiveness of \DeepGroundRepair compared to baseline methods, we assess the final DL programs from three complementary perspectives: static checking, dynamic checking, and semantic checking.

\paragraph{(1) Static Checking}
We employ Pylint~\cite{Pylintcodeanalysispythonwwwpylintorg}, a widely used static analysis tool, to identify API-level syntactic issues. We focus on the \textit{Error} messages reported by Pylint\footnote{We exclude other Pylint message categories (\textit{Fatal}, \textit{Refactor}, \textit{Warning}, and \textit{Convention}) because they are rare or capture lower-severity issues. We also filter out false positives due to TensorFlow's lazy-loading mechanism, such as the \texttt{import-error} report for \texttt{tensorflow.keras.layers}.}, which indicate severe issues that may affect program correctness. Based on their categories, we define two metrics: \textit{API Hallucination Count} and \textit{API Parameter Misuse Count}.

\textbullet\ \textbf{\textit{API Hallucination Count}} quantifies references to nonexistent APIs. It is computed as the total number of syntactic issues reported by Pylint, such as \texttt{no-member}, \texttt{import-error}, and \texttt{no-name-in-module}. 

\textbullet\ \textbf{\textit{API Parameter Misuse Count}} quantifies violations of API signatures. It is computed as the total number of syntactic issues reported by Pylint, such as \texttt{unexpected-keyword-arg}, \texttt{no-value-for-parameter}, \texttt{too-many-function-args}, and \texttt{redundant-keyword-arg}. 

\paragraph{(2) Dynamic Checking}
We execute each generated DL program and measure its \textit{Code Executability Rate} (\textit{CER}).

\textbullet\ \textbf{\textit{Code Executability Rate}} measures the proportion of generated programs that can be executed successfully without runtime errors. It reflects whether LLM-generated programs satisfy the basic usability requirements for DL tasks. Let $\mathcal{D}$ denote the set of generated programs and $\mathcal{D}_{\mathrm{exec\text{-}success}}$ denote the subset of executable programs. \textit{CER} is defined as:

\begin{equation}
\mathrm{\textit{CER}} = \frac{|\mathcal{D}_{\mathrm{exec\text{-}success}}|}{|\mathcal{D}|}.
\end{equation}

\paragraph{(3) Semantic Checking}
We introduce \textit{Semantic Consistency Rate} (\textit{SCR}) to evaluate whether repaired programs preserve the semantics of the original programs.

\textbf{\textit{Semantic Consistency Rate}} assesses whether repaired programs preserve the semantics of their pre-repair versions. However, exact assessment requires labor-intensive, fine-grained analysis of architectural changes, such as layer additions or removals. Therefore, we use effective-line-count preservation as a proxy for semantic consistency. Let $\mathcal{D}_{\mathrm{repair}}$ denote the set of programs subjected to repair and $\mathcal{D}_{\mathrm{preserved}}$ the subset of programs whose effective line counts remain unchanged. Specifically, \textit{SCR} is defined as:

\begin{equation}
  \mathrm{\textit{SCR}} =
  \frac{|\mathcal{D}_{\mathrm{preserved}}|}
  {|\mathcal{D}_{\mathrm{repaired}}|}.
\end{equation}

\subsubsection{Stage-specific Effectiveness Metrics} \label{sec:stage_effectiveness}
To evaluate whether each stage achieves its intended objective, we employ dedicated metrics tailored to the corresponding stage of \DeepGroundRepair.
\paragraph{(1) API Knowledge Retrieval.}
We evaluate Stage~1 by comparing the set of top-$K$ retrieved APIs with the ground-truth API set annotated for each task. Based on their overlap, we compute \textit{Precision@K} and \textit{Recall@K}. Let $G$ denote the ground-truth API set and $R^K$ the set of top-$K$ retrieved APIs.

\textbullet\ \textbf{\textit{Precision@K}} measures the proportion of relevant APIs among the top-$K$ retrieved APIs. The definition is as follows.
\begin{equation}
\mathrm{\textit{Precision@K}} = \frac{|R^K \cap G|}{K}.
\end{equation}

\textbullet\ \textbf{\textit{Recall@K}} measures the proportion of ground-truth APIs covered by the top-$K$ retrieved APIs. The definition is as follows.
\begin{equation}
\mathrm{\textit{Recall@K}} = \frac{|R^K \cap G|}{|G|}.
\end{equation}

\paragraph{(2) Root-Cause Identification.}
We evaluate Stage~2 using \textit{Root-Cause API Localization Accuracy}.

\textbullet\ \textbf{\textit{Root-Cause API Localization Accuracy}} measures the proportion of initial shape-mismatch failures for which Stage~2 correctly identifies the root cause API that is responsible for the mismatch. Let $\mathcal{D}_{\mathrm{shape\text{-}error}}$ denote the set of programs with initial shape-mismatch failures, and let $\mathcal{D}_{\mathrm{root\text{-}localized}}$ denote the subset for which the localized root-cause API matches the manually annotated root-cause API. We define \textit{RCLA} as:
\begin{equation}
\mathrm{\textit{RCLA}} =
\frac{|\mathcal{D}_{\mathrm{root\text{-}localized}}|}
     {|\mathcal{D}_{\mathrm{shape\text{-}error}}|}.
\end{equation}

\paragraph{(3) Shape-Aware Repair.}
We evaluate Stage~3 using \textit{Shape-Mismatch Error Repair Success Rate} (\textit{ShapeRSR}).

\textbullet\ \textbf{\textit{Shape-Mismatch Error Repair Success Rate}}
Following prior work~\cite{codecorrectness}, we adapt Repair Success Rate to measure successful fixes of shape-mismatch errors. Let $\mathcal{D}_{\mathrm{shape\text{-}error}}$ denote the programs that initially fail due to shape mismatches, and $\mathcal{D}_{\mathrm{shape\text{-}error}}^{\mathrm{repaired}}$ denote the successfully repaired subset. ShapeRSR is defined as:

\begin{equation}
  \mathrm{\textit{ShapeRSR}} =
  \frac{|\mathcal{D}_{\mathrm{shape\text{-}error}}^{\mathrm{repaired}}|}
       {|\mathcal{D}_{\mathrm{shape\text{-}error}}|}.
\end{equation}

\subsubsection{Token Efficiency.}
To evaluate token efficiency, we examine token cost alongside effectiveness. For each model--library configuration, token cost is measured as the average number of completion tokens across the 100 tasks.

For generation-only baselines (\textit{Baseline~1} and \textit{Baseline~2}):
\begin{equation}
\mathrm{\textit{Token Cost}} = T_{\mathrm{generation}}.
\end{equation}
 
For methods with a repair stage (\textit{Baseline~3}, \textit{Baseline~4}, and \DeepGroundRepair)\footnote{For \textit{Baselines~3 and~4}, $T_{\mathrm{generation}}$ is taken from \textit{Baselines~1 and~2}, respectively; for \DeepGroundRepair, it includes all completion tokens generated in Stage~1. If the initial code executes successfully, no repair is performed and $T_{\mathrm{repair}}=0$.}:
\begin{equation}
\mathrm{\textit{Token Cost}} = T_{\mathrm{generation}} + T_{\mathrm{repair}}.
\end{equation}
 
\subsection{Benchmark}
In this work, we adopt DeepEval~\cite{DeepEval} as our evaluation benchmark. To the best of our knowledge, DeepEval was the only benchmark specifically designed to evaluate model-level DL code generation when this study was conducted. It comprises 100 tasks, each consisting of a natural-language requirement and a reference implementation. The requirement specifies the model inputs, architecture, and outputs. Unlike general-purpose code generation benchmarks centered on standalone functions\cite{chenEvaluatingLargeLanguage,GoodBadMissingNeuralCodeGenerationMachineLearningTasks}, DeepEval focuses on the construction of complete DL models, enabling the evaluation of LLM-generated DL programs at the model level.

We evaluate all methods on the complete DeepEval benchmark. Its diversity in model architectures and task complexity provides a broad and consistent basis for comparing different methods. Moreover, the library-independent task descriptions allow the same tasks to be evaluated consistently across TensorFlow, PyTorch, and PaddlePaddle. For the retrieval evaluation in Section~\ref{sec:rq_stage1}, we derive library-specific ground-truth API sets by mapping APIs in the Keras reference implementations to their counterparts in TensorFlow, PyTorch, and PaddlePaddle.

\subsection{Implementation} \label{sec:implementation}
\subsubsection{API Document Database} 
To support retrieval in Stages~1 and~3 of \DeepGroundRepair, we construct library-specific API documentation databases for the three target DL libraries. We first use \textit{BeautifulSoup} to parse the official online API documentation of TensorFlow\footnote{\url{https://www.tensorflow.org/api_docs/python/tf}}, PyTorch\footnote{\url{https://docs.pytorch.org/docs/2.11/pytorch-api.html}}, and PaddlePaddle\footnote{\url{https://www.paddlepaddle.org.cn/documentation/docs/en/api/index_en.html}}, collecting 3,686, 4,137, and 1,531 API documents, respectively. We then preprocess the raw documents using GPT-5.4-mini to remove noise and shorten subsequent prompts, thereby helping the LLM focus on essential API information~\cite{liu2024lost}. Specifically, we normalize API signatures, remove nonessential parameters, and condense API and parameter descriptions. Finally, we organize the processed documents into library-specific databases and build separate sparse and dense retrieval indexes for each database.

\subsubsection{Retrieval Configuration}
Using the constructed API documentation databases, we configure both sparse and dense retrieval. For sparse retrieval, we use BM25~\cite{robertson2009probabilistic}, implemented with the \texttt{Whoosh} package~\cite{Whoosh}, to rank API documents based on lexical matches between each query and the structured API fields. For dense retrieval, we use BGE-M3~\cite{chen-etal-2024-m3} to encode queries and documents into embedding vectors. FAISS~\cite{Douze2026FAISS} indexes the document vectors and performs similarity search. In \DeepGroundRepair, existing and nonexistent API candidates are routed to sparse and dense retrieval, respectively. The top-ranked document for each query is retained, yielding up to 10 API documents as context for the LLM. By contrast, Hybrid-RAG applies both retrieval methods to each query and fuses their ranked results using reciprocal rank fusion (RRF)~\cite{Reciprocalrankfusion}, retaining the top 10 documents after fusion.

\subsubsection{Studied LLMs}\label{sec:studied_llms}
As detailed in Table~\ref{tab:LLMs}, we evaluate LLMs from the Gemma and Mistral families, covering three parameter-scale ranges within each family: below 10B, 10B--20B, and above 20B. This selection enables us to examine the generalizability of \DeepGroundRepair across model families and scales, as presented in Section~\ref{sec:rq4_generalizability}. All selected models are open-weight and publicly available on HuggingFace~\cite{Huggingface}, providing greater control over model deployment and facilitating transparent and reproducible evaluation.
\input{tables/studied_llms.tex}

\subsubsection{Environment}
To ensure a fair comparison, we evaluate all methods using identical hardware and software configurations. Specifically, all experiments are conducted on a platform equipped with two NVIDIA V100 GPUs and running CentOS Linux 7.8.2003 (Core), Python 3.10.0, CUDA 12.1, and cuDNN 8.9.5.29. Within this environment, we use TensorFlow 2.16.1, PyTorch 2.11.0, and PaddlePaddle 3.3.0, which were the latest releases available when the experiments were conducted. We run all LLMs at \texttt{bfloat16} precision. For decoding, we adapt nucleus sampling~\cite{curiouscaseneuraltextdegeneration} with top-$p$ set to 0.95 for all models. We set the temperature to 0.8 for Gemma models and 0.1 for Mistral models; the latter follows the decoding guidance in the official Mistral model cards.\footnote{\url{https://huggingface.co/mistralai/Ministral-3-3B-Instruct-2512}.} Finally, to ensure sufficient capacity for complete outputs, we set the maximum generation length to 2,000 tokens for draft code generation and 3,000 tokens for code repair.

\section{Experimental Results}
\subsection{RQ1: End-to-End Effectiveness} \label{sec:end_to_end_effectiveness}
To assess the end-to-end effectiveness of \DeepGroundRepair, we evaluate whether it alleviates the two problems identified in Section~\ref{sec:introduction}: API-level syntactic issues and runtime execution failures. We present the detailed results in the order of static checking, dynamic checking, and semantic checking, using the metrics defined in Section~\ref{sec:end_to_end_effectiveness_metrics}.

\subsubsection{Static checking}
We evaluate two API-level syntactic issues: API hallucination and API parameter misuse. As shown in Table~\ref{tab:rq1_sc_combined}, averaged over comparisons with four baselines across 18 model--library configurations, \DeepGroundRepair produces 2.14 fewer API hallucinations and 3.26 fewer API parameter misuses per 10 programs.

For API hallucination, \DeepGroundRepair achieves the lowest average \textit{API Hallucination Count} on TensorFlow and PaddlePaddle, with 0.10 and 3.33 issues per 10 program files, respectively. Compared with the four baselines, it reduces hallucinations by 0.25--0.71 issues on TensorFlow and 4.09--7.10 issues on PaddlePaddle. The particularly large reductions on PaddlePaddle suggest that \DeepGroundRepair is especially beneficial when LLMs have limited parametric knowledge of less widely used DL libraries. On PyTorch, \DeepGroundRepair outperforms \textit{Baselines~1--3} and remains close to \textit{Baseline~4}, with only 0.18 more issues per 10 program files.

For API parameter misuse, \DeepGroundRepair achieves the lowest average \textit{API Parameter Misuse Count} on TensorFlow, reducing the count to 0.84 and producing 2.79--4.51 fewer issues than the baselines. On PyTorch, all methods exhibit near-zero misuse counts, leaving little room for improvement. On PaddlePaddle, \DeepGroundRepair substantially outperforms \textit{Baseline~1} and \textit{Baseline~3}, remains comparable to \textit{Baseline~2}, and produces 0.29 more issues than \textit{Baseline 4}. Thus, the reductions are most pronounced in settings where the baselines exhibit substantial API misuse.

\input{tables/results/RQ1/RQ1_issues.tex}
\input{tables/results/RQ1/RQ1CER.tex}

\input{tables/results/RQ1/rq1_semantic_consistency_rate.tex}

\vspace{2em}
\subsubsection{Dynamic checking}
Static checking captures API-level syntactic issues but cannot reveal runtime failures; we therefore execute the final programs and measure their executability using \textit{CER}. As shown in Table~\ref{tab:rq1_cer}, averaged over the comparisons with four baselines across the 18 model--library configurations, \DeepGroundRepair achieves a relative \textit{CER} improvement of 85.3\%.

As shown in Table~\ref{tab:rq1_cer}, \DeepGroundRepair achieves the highest overall \textit{CER} on all three DL libraries, reaching 58.2\% on TensorFlow, 41.2\% on PyTorch, and 31.5\% on PaddlePaddle. Compared with the strongest baseline for each library, \DeepGroundRepair yields relative gains of 13.7\%, 31.6\%, and 60.0\%, respectively. Compared with the generation-only baselines, it improves \textit{CER} by 54.0\%--275.9\%, showing the benefit of applying targeted repair to initially generated code. It also outperforms the self-debugging baselines by 13.7\%--75.3\%, demonstrating that the complete \DeepGroundRepair pipeline is more effective than existing traceback-guided self-debugging methods.

Across 18 model--library configurations, \DeepGroundRepair achieves the highest \textit{CER} in 15 configurations and ties for the highest in one. The only two configurations in which it is outperformed are Mistral-14B and Mistral-24B on TensorFlow, where \textit{Baseline 4} achieves \textit{CER} of 66\% and 83\%, compared with 59\% and 66\% for \DeepGroundRepair. However, \textit{Baseline 4} exhibits lower \textit{SCR} in these configurations: 75.8\% and 83.0\%, compared with 98.0\% and 91.9\% for \DeepGroundRepair. These cases highlight the need to consider executability and semantic consistency jointly, as discussed below.

\subsubsection{Semantic checking}
To assess whether improved executability comes at the cost of reduced semantic consistency, we use \textit{SCR} to compare programs before and after repair. As shown in Table~\ref{tab:rq1_semantic_consistency_rate}, \DeepGroundRepair achieves average \textit{SCR} values of 94.02\%, 96.45\%, and 93.73\% on TensorFlow, PyTorch, and PaddlePaddle, respectively. At the configuration level, it obtains the highest \textit{SCR} in 16 of the 18 model--library configurations. Overall, these results represent an average relative improvement of 18.1\% over the two self-debugging baselines across the three libraries.

The importance of jointly considering \textit{CER} and \textit{SCR} is particularly evident for Mistral-14B and Mistral-24B on TensorFlow. These are the only two configurations in which \textit{Baseline~4} achieves a higher \textit{CER} than \DeepGroundRepair. However, its corresponding \textit{SCR} values are only 75.8\% and 83.0\%, compared with 98.0\% and 91.9\% for \DeepGroundRepair, respectively. These results indicate that higher executability can coincide with lower semantic consistency, underscoring the need to evaluate both metrics jointly.

\begin{tcolorbox}[
  colback=mycolorsummary!50!white,
  colframe=black!65,
  title=\textbf{Summary of Answers to RQ1},
  boxsep=1pt,
  left=6pt,
  right=6pt
]
Overall, \DeepGroundRepair demonstrates strong end-to-end effectiveness across all three evaluation dimensions. Averaged across the evaluated comparisons, it produces 2.14 fewer API hallucinations and 3.26 fewer API parameter misuses per 10 program files and achieves an average relative \textit{CER} improvement of 85.3\%. Compared with the self-debugging baselines, it also improves \textit{SCR} by an average of 18.1\%. These results show that \DeepGroundRepair substantially improves runtime executability while generally reducing API-level syntactic issues and achieving higher semantic consistency during repair.
\end{tcolorbox}

\subsection{RQ2: Ablation Study}
To quantify the contribution of each stage of \DeepGroundRepair, we conduct an ablation study in which one stage is removed at a time while all other stages remain unchanged. Following the evaluation procedure described in Section~\ref{sec:end_to_end_effectiveness}, we assess each ablated variant through static checking, dynamic checking, and semantic checking using the end-to-end effectiveness metrics defined in Section~\ref{sec:end_to_end_effectiveness_metrics}. The variants are described below.

\begin{itemize}
  \item \textit{Variant 1} (w/o \textit{Reasoning-Augmented Classification Retrieval}):
  Removes retrieved DL API knowledge and instead uses vanilla LLM generation to produce draft programs.
  \item \textit{Variant 2} (w/o \textit{Root-Cause Localization}):
  Removes the localized root-cause API and parameter masks from the repair context, using only the raw runtime traceback as repair guidance.
  \item \textit{Variant 3} (w/o \textit{Shape-A-Aware Self-Repair}):
  Removes tensor-shape annotations and shape-related API knowledge from the repair context, leaving the LLM without explicit shape guidance.
\end{itemize}

\input{tables/results/RQ2/rq2_issues.tex}
\textbf{Static checking.}
As shown in Table~\ref{tab:rq2_sc_combined}, removing \textit{Reasoning-Augmented Classification Retrieval} in \textit{Variant 1} consistently lowers static correctness across all three libraries. Compared with \DeepGroundRepair, \textit{Variant 1} increases \textit{API Hallucination Count} by 0.30--1.31 and \textit{API Parameter Misuse Count} by 0.05--5.92 issues per 10 program files. By contrast, \textit{Variant 2} and \textit{Variant 3} have smaller and less consistent effects. Although they occasionally reduce issue counts for individual library--metric combinations, these improvements are isolated and do not generalize across libraries. Overall, the consistent performance degradation of \textit{Variant 1} relative to \DeepGroundRepair demonstrates the importance of \textit{Reasoning-Augmented Classification Retrieval} in reducing API-level syntactic issues.

\input{tables/results/RQ2/RQ2_cer.tex}
\textbf{Dynamic checking.}
As shown in Table~\ref{tab:rq2_ab}, \textit{Variant 1} yields relative \textit{CER} decreases of 9.3\% on TensorFlow and 47.8\% on PaddlePaddle, despite an increase of 4.5\% on PyTorch. \textit{Variant 2} exhibits a similar library-dependent pattern, decreasing \textit{CER} by 20.5\% on PyTorch and 18.8\% on PaddlePaddle while producing a marginal increase of 0.8\% on TensorFlow. Although their effects vary across libraries, removing either API knowledge grounding or root-cause localization generally reduces code executability. By contrast, \textit{Variant 3} produces the largest and most consistent degradation, decreasing \textit{CER} by 21.2\%--39.8\% across all three libraries. Its average relative decrease reaches 32.9\%, substantially exceeding the decreases of 17.5\% and 12.8\% observed for \textit{Variant 1} and \textit{Variant 2}, respectively. Taken together, these results demonstrate that \textit{Shape-Aware Self-Repair} makes the strongest contribution to improving code executability.

\input{tables/results/RQ2/rq2_semantic_consistency_rate.tex}
\textbf{Semantic consistency.}
As shown in Table~\ref{tab:rq2_semantic_consistency_rate}, \textit{Variant 1} consistently reduces \textit{SCR}, with relative decreases of 2.1\%--8.7\% across the three libraries. \textit{Variant 3} has a limited and mixed effect, resulting in an average relative decrease of only 2.4\%. By contrast, \textit{Variant 2} produces the largest degradation, reducing \textit{SCR} by 8.7\%--21.6\% across the three libraries, or 14.6\% on average. Overall, the substantial performance degradation of \textit{Variant 2} relative to \DeepGroundRepair demonstrates the importance of \textit{Root-Cause Localization}. By directing repairs toward the relevant faulty APIs, this stage helps preserve semantic consistency.

\begin{tcolorbox}[
  colback=mycolorsummary!50!white,
  colframe=black!65,
  title=\textbf{Summary of Answers to RQ2},
  boxsep=1pt,
  left=6pt,
  right=6pt
]
Overall, the ablation results demonstrate that the three stages make complementary contributions to the performance of \DeepGroundRepair. Stage~1 is the primary contributor to reducing API-level syntactic issues and also improves executability overall. Stage~2 makes the largest contribution to preserving semantic consistency while supporting executability by directing repairs toward localized root-cause APIs. Stage~3 makes the strongest and most consistent contribution to code executability while largely preserving semantic consistency.
\end{tcolorbox}

\subsection{RQ3: Stage-Specific Analysis}
In this section, we evaluate how effectively the three stages of \DeepGroundRepair address the corresponding challenges identified in Section~\ref{sec:introduction}: retrieving task-relevant API knowledge, localizing root-cause APIs, and repairing incorrect parameters under tensor-shape constraints. This analysis further clarifies the mechanisms underlying each stage's contribution and helps explain the ablation results in RQ2.

\subsubsection{Stage~1: Reasoning-Augmented Classification Retrieval} \label{sec:rq_stage1}
We evaluate the retrieval accuracy of \textit{Reasoning-Augmented Classification Retrieval} of \DeepGroundRepair using \textit{Precision@10} and \textit{Recall@10} and compare it with the Hybrid-RAG.

\textbf{API ground-truth set construction.}
To establish relevance labels for evaluating retrieval accuracy, we construct a library-specific API ground-truth set for each task based on the reference implementation of DeepEval. Specifically, one author extracts API calls from the reference code and maps them to corresponding or functionally equivalent APIs in TensorFlow, PyTorch, and PaddlePaddle. Another author then independently verifies all extracted APIs and mappings against the official library documentation to ensure their correctness and equivalence. The resulting task-specific sets are used as relevance labels for computing \textit{Precision@10} and \textit{Recall@10}. 

\vspace{-0.3em}

\input{tables/results/RQ1/RQ1_RA.tex} 
\textbf{Overall result.}
As shown in Table~\ref{tab:rq3_ra}, Stage~1 of \DeepGroundRepair substantially outperforms Hybrid-RAG on both metrics across all 18 combinations of recommender LLMs and DL libraries. Averaged across the six LLMs, the improvements range from 55.4 to 74.7 percentage points in \textit{Precision@10} and from 56.2 to 83.5 percentage points in \textit{Recall@10} across the three libraries. These substantial and consistent improvements demonstrate the effectiveness of \textit{Reasoning-Augmented Classification Retrieval} in retrieving task-relevant APIs.

\input{tables/method/stage1_example.tex}
\textbf{Mechanism analysis.} 
\textit{Reasoning-Augmented Classification Retrieval} verifies each LLM-recommended API and routes it according to its existence: existing APIs to sparse retrieval and hallucinated candidates to dense retrieval. This strategy preserves the precise lexical signals of existing API names while exploiting the semantic information in hallucinated candidates to recover valid alternatives. As illustrated in Table~\ref{tab:racr_example}, \texttt{torch.nn.functional.add} does not exist but captures the intended tensor-addition operation. \textit{Reasoning-Augmented Classification Retrieval} therefore uses dense retrieval rather than sparse retrieval and successfully recovers the valid API \texttt{torch.add}.

\subsubsection{Stage~2: Root-Cause Localization}
Stage~2 aims to localize the root-cause API underlying an execution failure rather than merely the API at which the failure manifests. On programs that fail during initial execution, we use two complementary metrics to assess failure-site--root-cause discrepancies and the localization accuracy of Stage~2. \textit{Failure-Site--Root-Cause Discrepancy Rate} (\textit{FRDR}) measures the proportion of cases where the traceback-reported failure site is not the actual root-cause location. \textit{Root-Cause API Localization Accuracy} (\textit{RCLA}) measures the proportion of cases in which Stage~2 correctly identifies the manually annotated root-cause API.
 
\textbf{Root-cause API ground-truth set construction.}
To balance representativeness and annotation cost, we used the medium-sized model from each family, Gemma-12B and Mistral-14B. From the 351 programs generated by these models that failed during initial execution, we selected 120 cases through stratified random sampling~\cite{neuendorf2017content} by model and DL library. One author manually annotated the root-cause API of each failure based on the source code and runtime traceback, without access to the predictions of Stage~2. A second author reviewed ambiguous cases to finalize the ground-truth labels. These labels were then used to evaluate the root-cause APIs localized by Stage~2.

\vspace{-0.4em}
\input{tables/results/RQ3/rq3_api_acc.tex}
\vspace{-0.4em}

\textbf{Overall result.}
As shown in Table~\ref{tab:rq3_api_localization}, across all 120 cases, the overall \textit{FRDR} is 94.2\%, indicating that the reported failure site is generally a poor proxy for the underlying cause. Nevertheless, Stage~2 achieves an overall \textit{RCLA} of 82.5\%, with library-level accuracy ranging narrowly from 80.9\% to 83.7\%.

\textbf{Mechanism analysis.}
The high \textit{FRDR} indicates that a tensor incompatibility is usually reported downstream of the API that caused it, making traceback information alone insufficient for reliable root-cause localization. Stage~2 instead backtracks tensor-transformation dependencies from the reported failure site to upstream API configurations, enabling consistent localization accuracy across libraries. This effect is particularly evident for TensorFlow: despite an \textit{FRDR} of 100.0\%, Stage~2 achieves an \textit{RCLA} of 83.3\%.

\subsubsection{Stage~3: Shape-Aware Self-Repair}
To isolate the contribution of explicit tensor-shape guidance in Stage~3, we evaluate programs that initially fail due to shape mismatches. We compare \DeepGroundRepair with \textit{Variant~3}, which uses identical draft programs and Stage~2 localization outputs but omits shape-aware guidance from Stage~3.
 
\vspace{-0.4em}

\input{tables/results/RQ3/rq3_shape_rsr.tex} 
\vspace{-0.4em}
\input{figures/plot/RQ3/sankey.tex}

\textbf{Overall result.}
Table~\ref{tab:rq3_shape_rsr} shows that explicit tensor-shape guidance generally improves \textit{ShapeRSR} for programs with initial tensor-shape mismatch failures. Across the evaluated model--library configurations, \DeepGroundRepair generally outperforms or matches \textit{Variant~3}, with degradation occurring only occasionally. These results indicate that shape-aware repair effectively guides the LLM toward parameter changes that restore shape consistency. Nevertheless, some programs still fail after repair because resolving the initial shape mismatch may expose subsequent execution failures. We further analyze these remaining failures and their causes below.

\textbf{Mechanism analysis.}
Figure~\ref{fig:root_cause_transition_sankey} shows how programs with initial intra-path and inter-path mismatches transition to different post-repair outcomes under \textit{Variant~3} and \DeepGroundRepair. Across both initial mismatch types, the \textit{Fixed} outcome contains 6 cases under \textit{Variant~3} versus 31 under \DeepGroundRepair in TensorFlow, 15 versus 50 in PyTorch, and 10 versus 38 in PaddlePaddle.

For initial intra-path mismatches, the improvement is particularly pronounced in PyTorch and PaddlePaddle. Among the 91 initial intra-path cases in PyTorch, the flow to \textit{Fixed} contains 15 cases under \textit{Variant~3}, compared with 50 under \DeepGroundRepair. This difference suggests that explicit shape information helps the LLM produce parameter configurations that restore shape compatibility along a single execution path.

For initial inter-path mismatches, the outcome flows differ across libraries. In TensorFlow, more cases flow directly from \textit{Inter-path} to \textit{Fixed} under \DeepGroundRepair. In PyTorch and PaddlePaddle, more cases instead flow from \textit{Inter-path} to the post-repair \textit{Intra-path} outcome. Specifically, among the 30 initial inter-path cases in PyTorch, this flow contains 3 cases under \textit{Variant~3} versus 13 under \DeepGroundRepair. Nevertheless, the transition indicates that the original inter-path incompatibility was resolved, allowing execution to expose a subsequent intra-path mismatch.

\begin{tcolorbox}[
  colback=mycolorsummary!50!white,
  colframe=black!65,
  title=\textbf{Summary of Answers to RQ3},
  boxsep=1pt,
  left=6pt,
  right=6pt
]
Overall, the RQ3 results show that each stage of \DeepGroundRepair effectively addresses its corresponding challenge. Stage~1 improves task-relevant API knowledge retrieval, Stage~2 accurately localizes upstream root-cause APIs beyond traceback-reported failure sites, and Stage~3 generally improves tensor-shape mismatch repair. Together, these findings explain the complementary stage contributions observed in RQ2 and provide stage-level evidence for the end-to-end effectiveness reported in RQ1.
\end{tcolorbox}

\subsection{RQ4: Generalizability} \label{sec:rq4_generalizability}
To answer RQ4, we assess the generalizability of \DeepGroundRepair by examining whether its effectiveness remains consistent across experimental configurations. Using the RQ1 results, we group the 18 model--library configurations along three dimensions: model family, model scale, and DL library. Figure~\ref{fig:generalizability_matrix} compares the group-level performance of \DeepGroundRepair with the corresponding baseline averages, which are aggregated from the average baseline performance of each configuration.

\input{figures/plot/RQ4/RQ4.tex}

\subsubsection{Model Families}
We first present the results for the Gemma and Mistral families.

\textbf{Comparison across model families.}
The two model families differ noticeably in their absolute performance. With \DeepGroundRepair, Mistral produces 0.96 syntactic issues per 10 program files, compared with 4.58 for Gemma. Mistral also attains a higher \textit{CER} of 48.8\%, whereas Gemma reaches 38.4\%. In contrast, \textit{SCR} remains comparably high for both families, reaching 93.9\% for Mistral and 95.6\% for Gemma. These differences indicate that model family remains an important source of variation in the quality of the resulting programs.

\textbf{Consistency of improvements.}
More importantly, the left-column panels of Figure~\ref{fig:generalizability_matrix} show that \DeepGroundRepair consistently outperforms the baseline averages for both families. It reduces syntactic issues from 12.22 to 4.58 for Gemma and from 4.13 to 0.96 for Mistral (top panel). It also increases \textit{CER} from 20.2\% to 38.4\% and from 34.8\% to 48.8\% (middle panel), and \textit{SCR} from 86.1\% to 95.6\% and from 74.4\% to 93.9\% (bottom panel), respectively. Thus, despite their different absolute performance, \DeepGroundRepair provides consistent improvements across both model families.

\subsubsection{Model Scales}
For the following comparison, we denote the three parameter-scale ranges defined in Section~\ref{sec:studied_llms} as small (below 10B), medium (10B--20B), and large (above 20B).

\textbf{Comparison across model scales.}
Across these evaluated scale groups, models with more parameters exhibit fewer syntactic issues and higher \textit{CER}. The small-model group produces 5.51 syntactic issues per 10 program files, whereas the medium- and large-model groups produce 2.06 and 0.73, respectively. Across the same groups, \textit{CER} rises from 31.0\% to 45.8\% and then to 54.0\%. By comparison, \textit{SCR} varies only between 93.9\% and 95.6\%. These results suggest that model scale is more clearly associated with syntactic correctness and code executability.

\textbf{Consistency of improvements.}
As shown in the model-scale column of Figure~\ref{fig:generalizability_matrix}, \DeepGroundRepair consistently outperforms the corresponding baseline averages across all three scale groups. For the small-, medium-, and large-model groups, \DeepGroundRepair achieves \textit{CER} values of 31.0\%, 45.8\%, and 54.0\%, respectively, compared with baseline averages of 14.1\%, 30.0\%, and 38.4\%. Its corresponding \textit{SCR} values are 94.6\%, 93.9\%, and 95.6\%, exceeding the baseline averages of 73.0\%, 83.3\%, and 84.5\%. Meanwhile, its syntactic-issue counts are 5.51, 2.06, and 0.73 per 10 program files, lower than the baseline averages of 16.32, 4.45, and 3.75. Taken together, these findings demonstrate consistent gains across all evaluated model-scale groups.

\subsubsection{DL Libraries}
Finally, we examine the results across TensorFlow, PyTorch, and PaddlePaddle.

\textbf{Comparison across DL libraries.}
The performance of \DeepGroundRepair varies across the target DL libraries. On TensorFlow, it achieves the fewest syntactic issues, at 0.94 per 10 program files, and the highest \textit{CER}, at 58.2\%. PyTorch ranks second, with 1.30 issues and a \textit{CER} of 41.2\%, whereas PaddlePaddle is the most challenging, with 6.08 issues and a \textit{CER} of 31.5\%. By contrast, \textit{SCR} remains stable across the three libraries, ranging from 93.7\% to 96.5\%.

\textbf{Consistency of improvements.}
Despite variations in absolute performance, \DeepGroundRepair consistently outperforms the corresponding baseline averages across all three libraries. On TensorFlow, PyTorch, and PaddlePaddle, it achieves \textit{CER} values of 58.2\%, 41.2\%, and 31.5\%, respectively, compared with baseline averages of 43.7\%, 24.4\%, and 14.5\%. Its corresponding \textit{SCR} values are 94.0\%, 96.5\%, and 93.7\%, exceeding the baseline averages of 80.9\%, 81.6\%, and 78.2\%. It also reduces the syntactic-issue counts to 0.94, 1.30, and 6.08 per 10 program files, compared with baseline averages of 4.98, 1.86, and 17.68. The particularly large reduction on PaddlePaddle indicates that \DeepGroundRepair remains effective in the most challenging API ecosystem.

\begin{tcolorbox}[
  colback=mycolorsummary!50!white,
  colframe=black!65,
  title=\textbf{Summary of Answers to RQ4},
  boxsep=1pt,
  left=6pt,
  right=6pt
]
Overall, the RQ4 results support the generalizability of \DeepGroundRepair across the evaluated settings. Across both model families, the three model-scale groups, and the three DL libraries, it consistently reduces API-level syntactic issues while improving code executability and semantic consistency relative to the baseline averages.
\end{tcolorbox}

\vspace{1em}
\subsection{RQ5: Token Efficiency}
To assess the token efficiency of \DeepGroundRepair, we jointly analyze its code executability and token cost. We use code executability, measured by \textit{CER}, as the effectiveness criterion because successful execution is a fundamental requirement for generated DL programs. Table~\ref{tab:rq5_token_cost_b3b4_conditional} reports the detailed \textit{Token Cost} across models, DL libraries, and methods, while Figure~\ref{fig:token_cost_vs_cer} visualizes the relationship between \textit{Token Cost} and \textit{CER}.

\input{tables/results/RQ5/RQ5_tokens.tex}
Compared with \textit{Baselines~1 and~2}, \DeepGroundRepair incurs 25.0\%--63.0\% higher \textit{Token Cost} because it additionally repairs failed programs. This additional cost is accompanied by relative \textit{CER} improvements of 54.0\%--275.9\%, as reported in RQ1. Because \textit{Baselines~3 and~4} also involve both generation and repair, they provide a more comparable basis for evaluating token efficiency. Relative to these baselines, \DeepGroundRepair reduces \textit{Token Cost} by 18.5\%--33.9\% while improving \textit{CER} by 13.7\%--75.3\%, thereby achieving both lower token cost and higher code executability than the self-debugging baselines.

\input{figures/plot/RQ5/RQ5.tex}
As shown in Figure~\ref{fig:token_cost_vs_cer}, the \textit{Token Cost} and \textit{CER} of each method are averaged across the six evaluated models and presented separately for the three DL libraries. Across all three DL libraries, \DeepGroundRepair exhibits a consistent pattern: it achieves substantially higher \textit{CER} than \textit{Baselines~1 and~2} at an additional token cost, while outperforming \textit{Baselines~3 and~4} in \textit{CER} with lower token cost. Overall, these results demonstrate that \DeepGroundRepair achieves a favorable trade-off between code executability and token cost across the evaluated settings.

\begin{tcolorbox}[
  colback=mycolorsummary!50!white,
  colframe=black!65,
  title=\textbf{Summary of Answers to RQ5},
  boxsep=1pt,
  left=6pt,
  right=6pt
]
Overall, \DeepGroundRepair demonstrates favorable token efficiency. Although it has a higher \textit{Token Cost} than the generation-only baselines, this additional cost is accompanied by substantially higher \textit{CER}. Compared with the self-debugging baselines, it reduces \textit{Token Cost} by 18.5\%--33.9\% while also achieving higher \textit{CER}, indicating a more favorable trade-off between effectiveness and token cost.
\end{tcolorbox}

\section{Discussion}
In this section, we discuss the limitations and opportunities of our approach, followed by threats to validity.
\subsection{Limitations of \DeepGroundRepair}
Although \DeepGroundRepair is effective in improving LLM-generated DL programs through API-grounded generation, root-cause localization, and shape-aware repair, it has several limitations.

First, \DeepGroundRepair primarily addresses API-level syntactic issues and tensor-shape mismatches, which are among the most prevalent failure categories observed in our evaluation. Its current design provides limited support for other DL-specific failures, including dtype or device incompatibilities, numerical instability, and training divergence. Incorporating failure-specific diagnostic capabilities and domain knowledge could enable more comprehensive analysis and repair of DL programs.

Second, the effectiveness of shape-aware analysis depends on observable tensor propagation and formalizable API semantics. Complex control flows, custom operators, and incomplete API documentation can hinder accurate tensor-flow tracing and shape-constraint validation. Advanced program-analysis techniques, such as symbolic execution, could strengthen tensor reasoning in these complex settings.

Finally, the current strategy performs parameter-level repair within a single iteration. Programs containing multiple independent errors may require repeated execution--repair cycles. Errors requiring architectural changes fall outside the scope of parameter-level repair and demand broader code revisions. Adaptive multi-step strategies could address these cases by dynamically determining the appropriate repair granularity and leveraging intermediate validation feedback. Such strategies could progressively resolve complex failures while maintaining repair efficiency.

\subsection{Threats to Validity}
We discuss potential threats to the internal and external validity of this study.

\textit{\textbf{Internal Validity.}}
The inherent stochasticity of LLMs may cause variations in generated programs. To reduce this variation, we use consistent model configurations, prompts, and experimental environments within each comparison. Metric selection may also affect the results. Although we employ complementary static, dynamic, and semantic checks, each has inherent limitations. Pylint detects only predefined issue categories and may produce false positives, although we exclude most identifiable cases. Successful execution does not guarantee functional correctness, and effective-line-count preservation is only a proxy for semantic consistency. Combining these metrics reduces reliance on any single measure. Finally, the ground-truth API sets and root-cause APIs used in the stage-specific evaluations require human annotation. We mitigate potential annotation bias by applying consistent guidelines and cross-checking annotations against reference implementations, API documentation, and runtime traces.

\textit{\textbf{External Validity.}}
The generalizability of our findings is limited by the scope of the evaluated tasks, LLMs, and DL libraries. We evaluate 100 model-level DL programming tasks from DeepEval, six open-weight LLMs from two model families, and three DL libraries: TensorFlow, PyTorch, and PaddlePaddle. Although these settings encompass diverse model architectures and library ecosystems, the results may not generalize to other LLMs, DL libraries, or large-scale real-world projects. Moreover, our evaluation focuses on API-related syntactic issues and tensor-shape mismatches. Other failure types, such as numerical instability and training-related issues, lie outside its scope. Evaluations spanning a broader range of DL tasks, LLMs, libraries, real-world projects, and failure scenarios are therefore needed to assess generalizability beyond the studied settings.

\section{Related Work}
\subsection{Bugs and Constraints in Deep Learning Programs}
DL programs define model architectures by composing library-specific APIs into computation graphs, in which tensors are propagated and transformed across operations~\cite{Tutorialdirectedacyclicgraphs, ExploringkerasfunctionalAPIcomplexmodelarchitecturesrutasoftware}. Prior studies have consequently focused on two aspects: common bugs in DL programs and the constraints governing API usage and tensor transformations.

\paragraph{Bugs in Deep Learning Programs}
Liang et al.~\cite{gDefects4DL} studied 64 real-world bugs in TensorFlow, PyTorch, and Keras programs and identified API misuse, tensor-shape mismatch, and type mismatch as common bug categories. Yang et al.~\cite{MisuAPI} further found that DL API misuses differ from those of general-purpose APIs, frequently involving incorrect parameter types and missing API calls. More recently, Ma et al.~\cite{DeepEval} evaluated LLM-generated DL programs and reported frequent API hallucinations and tensor-shape mismatches. The latter accounted for an average of 77.34\% of execution failures. Together, these findings show that reliable DL code generation requires correct API usage and consistent tensor shapes across operations.
 
\paragraph{Constraints in Deep Learning Programs}
Beyond characterizing DL program bugs, recent studies have examined the ability of LLMs to extract and reason about API constraints. Xie et al.~\cite{xie2025effective} evaluated LLMs for extracting formal specifications from comments and API documentation across TensorFlow, PyTorch, and MXNet. They considered data-type, structural, shape, and valid-value specifications. Their failure analysis identified \emph{missing domain knowledge} as a major source of extraction errors. Focusing on numerical constraints in PyTorch and NumPy APIs, Deng et al.~\cite{Deng2024CanLLMs} further showed that LLM performance degrades when handling uncommon inputs and complex numerical relationships in APIs such as \texttt{Conv2d}, \texttt{reshape}, and \texttt{Fold}. This finding indicates that LLMs have limited robustness in reasoning about implicit numerical and shape constraints. Together, these observations motivate the use of explicit API knowledge and tensor-shape information when repairing DL programs.

\subsection{Retrieval-Augmented Code Generation}
Retrieval-augmented generation (RAG) improves code generation by incorporating external knowledge into the generation context~\cite{Survey, wang-etal-2025-coderag, yang2025empirical}. Based on the retrieved knowledge source, existing approaches can be broadly divided into API documentation retrieval and relevant code retrieval.

\paragraph{API Documentation Retrieval.}
API documentation provides explicit knowledge of API functionality, signatures, parameters, and usage. Chen et al.~\cite{WhenLLMsmeetAPIdocumentation} showed that documentation retrieval improves code generation, particularly for less commonly used libraries. Zhou et al.~\cite{DocPromptinggeneratingcoderetrievingdocs} proposed DocPrompting, which retrieves relevant documentation using sparse and dense retrieval. Jain et al.~\cite{mitigatingcodeLLMhallucinationsAPIdocumentation} showed that documentation augmentation can reduce low-frequency API hallucinations and proposed DAG++, which selectively retrieves documentation based on API existence and model confidence. These studies establish the value of API documentation for grounding code generation, but do not explicitly address the gap between high-level task requirements and the fine-grained APIs required for DL implementation. \DeepGroundRepair addresses this gap by reasoning about task-relevant APIs before retrieval and routing candidates according to API existence.

\paragraph{Relevant Code Retrieval.}
Another line of work retrieves relevant code or repository context to provide implementation patterns. Gu et al.~\cite{Whatretrieveeffectiveretrieval} showed that relevant file context and invoked-API information improve repository-level code generation, whereas irrelevant context can degrade performance. Pan et al.~\cite{CATCODERRepository} combined sparse and dense code retrieval with statically extracted type context, while Li et al.~\cite{AceCoder} retrieved similar programs as in-context examples and generated intermediate artifacts to clarify requirements. These approaches primarily retrieve implementation context and reusable code patterns to guide generation. In contrast, \DeepGroundRepair retrieves library-specific API specifications to ground API usage during generation and support tensor-shape reasoning during repair.

\subsection{Repair of LLM-Generated Code}
Existing approaches to repairing LLM-generated code can be broadly divided into training-based and feedback-guided methods. The former learn repair behavior through model optimization, whereas the latter use program feedback to guide repair at inference time.

\paragraph{Training-Based Code Repair}
One line of work enhances code-repair capabilities through task-specific training on buggy programs and their corrected versions. Ding et al.~\cite{Ding2024CYCLE} proposed CYCLE, which constructs refinement examples from model-generated faulty code, test feedback, and canonical solutions for iterative self-refinement. Jiang et al.~\cite{jiang2024ledex} proposed LEDEX, which learns from explanation and refinement trajectories validated through execution, using supervised fine-tuning and reinforcement learning. Huang et al.~\cite{Huang2023FineTuningAPR} studied how code representations, fault-localization markers, and patch-generation strategies affect fine-tuned program repair. These approaches improve repair performance through dedicated training, whereas \DeepGroundRepair performs repair at inference time without model updates.

\paragraph{Feedback-Guided Code Repair}
Another line of work uses behavioral information from generated programs to guide post-generation refinement. Chen et al.~\cite{Teachinglargelanguagemodelsselfdebug} proposed Self-Debugging, which prompts an LLM to iteratively inspect and revise its code using test results or self-generated explanations. Zhang et al.~\cite{Selfedit} proposed Self-Edit, which summarizes failures observed from example tests as fault-aware comments for a separately trained code editor to revise the program. Shi et al.~\cite{codecorrectness} proposed MGDebugger, which structures repair according to program granularity by decomposing a program into hierarchical subfunctions, deriving local tests, and recursively repairing them from lower to higher levels. These methods demonstrate the effectiveness of program feedback for code repair. However, for DL programs, generic failure feedback provides limited guidance for identifying the upstream API configurations and tensor-shape dependencies underlying DL-specific failures. \DeepGroundRepair instead combines root-cause localization with explicit tensor-shape guidance for parameter-level repair.

\section{Conclusion and Future Work}
This paper presents \DeepGroundRepair, a three-stage framework for improving the syntactic correctness and runtime executability of LLM-generated DL programs. It targets API-level syntactic issues arising during code generation and tensor-shape mismatches that cause runtime failures. The framework progressively narrows the generation and repair spaces by retrieving task-relevant DL API knowledge, localizing the root-cause APIs of shape mismatches through static and dynamic analysis, and inferring shape-consistent API parameters from runtime tensor shapes.

We evaluate \DeepGroundRepair on DeepEval using six open-weight LLMs across TensorFlow, PyTorch, and PaddlePaddle. Across comparisons with the four baselines, \DeepGroundRepair substantially improves \textit{CER}, achieving an average relative gain of 85.3\%. It also reduces API hallucinations and API parameter misuses by 2.14 and 3.26 issues per 10 program files on average, respectively. For repaired programs, it improves \textit{SCR} by an average of 18.1\% relative to the two self-debugging baselines. Together, these findings demonstrate the effectiveness of the generation--localization--repair design of \DeepGroundRepair in improving the reliability of LLM-generated DL programs.

Future work will extend this generation--localization--repair paradigm to provide proactive and comprehensive support for DL program generation and repair. Integrating DL-specific constraints and lightweight verification into the generation process could help prevent failures before post-generation repair is required. Advanced program analyses could handle complex tensor flows. Adaptive multi-step repair strategies could address programs containing multiple interacting faults. Finally, we plan to expand diagnosis and repair beyond API-level issues and tensor-shape mismatches. Potential targets include dtype and device incompatibilities, numerical instability, and failures during training and deployment. Addressing these failures would provide broader support across the DL software lifecycle.

\begin{acks}
This work was supported by the National Natural Science Foundation of China under Grants 62372021, the National Key R\&D Program of China under Grant 2024YFB3311503, the State Key Laboratory of Complex \& Critical Software Environment under Grant SKLCCSE-2025ZX-02, and the Fundamental Research Funds for Beijing Municipal Universities under Grant 312000546325001.
\end{acks}


\bibliographystyle{ACM-Reference-Format}
\bibliography{acmart}

\appendix \label{sec:appendix}
\section{Prompting Templates}
This appendix presents the prompt templates used by \DeepGroundRepair. To ensure a fair comparison, prompt components unrelated to method-specific designs are kept identical across \DeepGroundRepair and the baselines. The complete prompts for all methods are available in our open-source repository.

\subsection{Stage~1: \textit{Reasoning-Augmented Classification Retrieval}}
As described in Section~\ref{sec:stage1}, Stage~1 involves two LLM interactions for API recommendation and code generation. The former recommends candidate API names from the task requirement for subsequent verification and retrieval, whereas the latter generates a DL program grounded in the retrieved API documentation.

\subsubsection{API Recommendation}
As shown in Figure~\ref{fig:prompt1}, the prompt combines the DeepEval task requirement, recommendation instructions, and an output format to elicit a structured list of candidate API names.

\input{prompts/stage1_1.tex}

\subsubsection{Draft Code Generation}
As shown in Figure~\ref{fig:prompt2}, the prompt combines the task requirement, generation instructions, an output format, and retrieved documentation containing the names, signatures, and descriptions of relevant APIs. These components guide the LLM in generating a complete model-level DL program.

\input{prompts/stage1_2.tex}

\subsection{Stage~3: \textit{Shape-Aware Self-Repair}}
Stage~3 uses separate prompt templates for the intra-path and inter-path repair strategies described in Section~\ref{sec:stage3}. The former provides local tensor-shape context, whereas the latter additionally supplies shape-related API knowledge for satisfying cross-path constraints.

\subsubsection{Intra-path Mismatch Repair}
As shown in Figure~\ref{fig:prompt3}, the prompt includes shape-annotated and parameter-masked code, the localized buggy API call, repair instructions, and an output format. The shape annotations provide the local tensor-shape constraints needed to infer the masked parameter values, while the repair instructions restrict changes to the localized buggy API call.

\input{prompts/stage3_1.tex}

\subsubsection{Inter-path Mismatch Repair}
As shown in Figure~\ref{fig:prompt4}, the prompt combines shape-annotated and parameter-masked code, cross-path shape constraints, retrieved API documentation, repair instructions, and an output format. The documentation describes the masked parameters and provides the shape-transformation formula of the buggy API, guiding the LLM to infer parameter values that satisfy the cross-path constraints.

\input{prompts/stage3_2.tex}

\end{document}

%% file: figures/motivation/code_syntax/syntactic_issues.tex
\begin{figure}[htbp]
    \centering
    \includegraphics[width=0.78\textwidth]{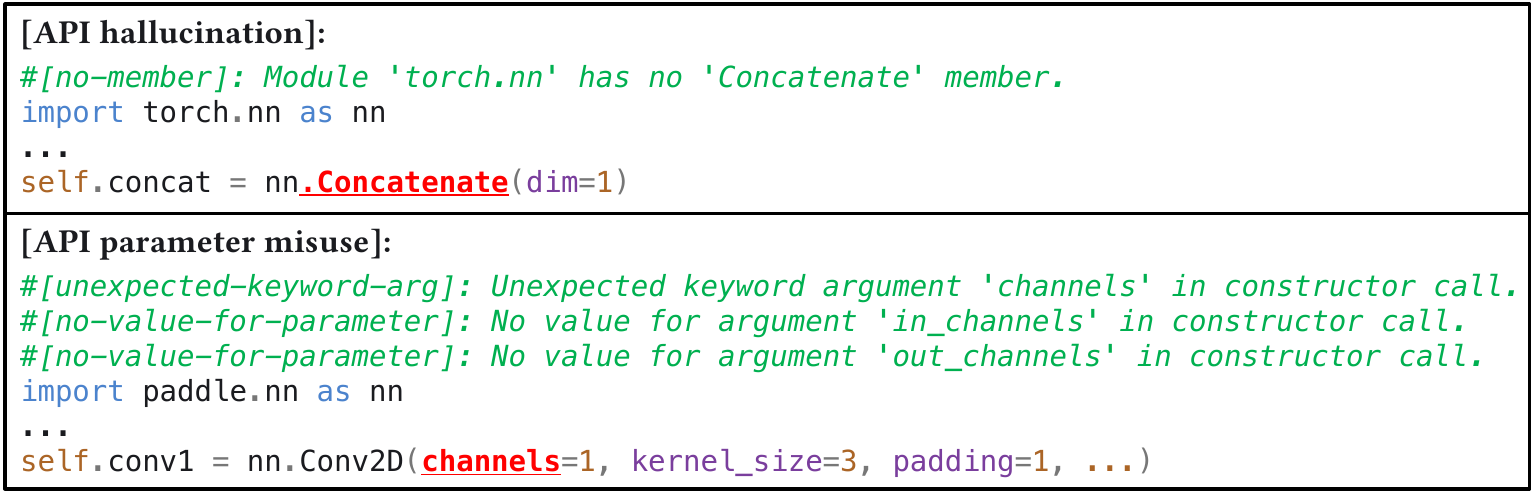}
    \caption{Examples of API-level syntactic issues in DL programs generated by the vanilla Gemma-12B model.}
    \label{fig:syntactic_issues_label}
\end{figure} 

%% file: figures/motivation/code_syntax/retrieval_failure.tex
\begin{figure}[htbp]
\centering
\includegraphics[width=0.92\textwidth]{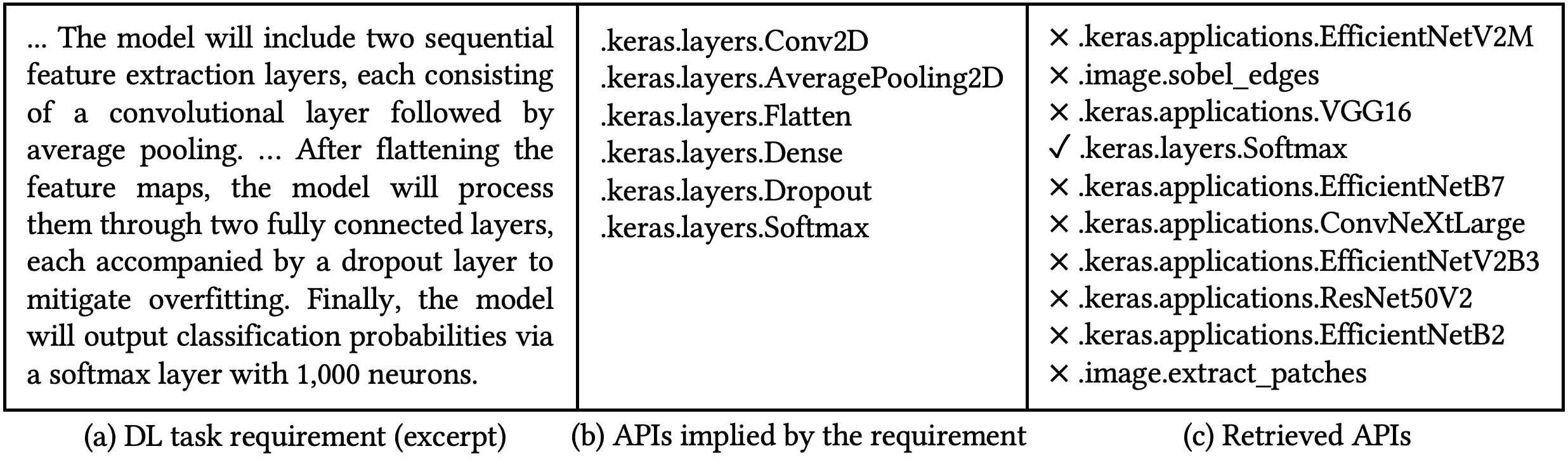}
\caption{Example of API retrieval failure, where only one of six required tensorflow APIs appears among the top 10 results. The \texttt{tensorflow.} prefix is omitted from the API names for brevity.}
\label{fig:retrieval_failure}
\end{figure}

%% file: figures/motivation/code_executability/executability_failure.tex
\begin{figure}[h]
    \centering
    \includegraphics[width=0.65\textwidth]{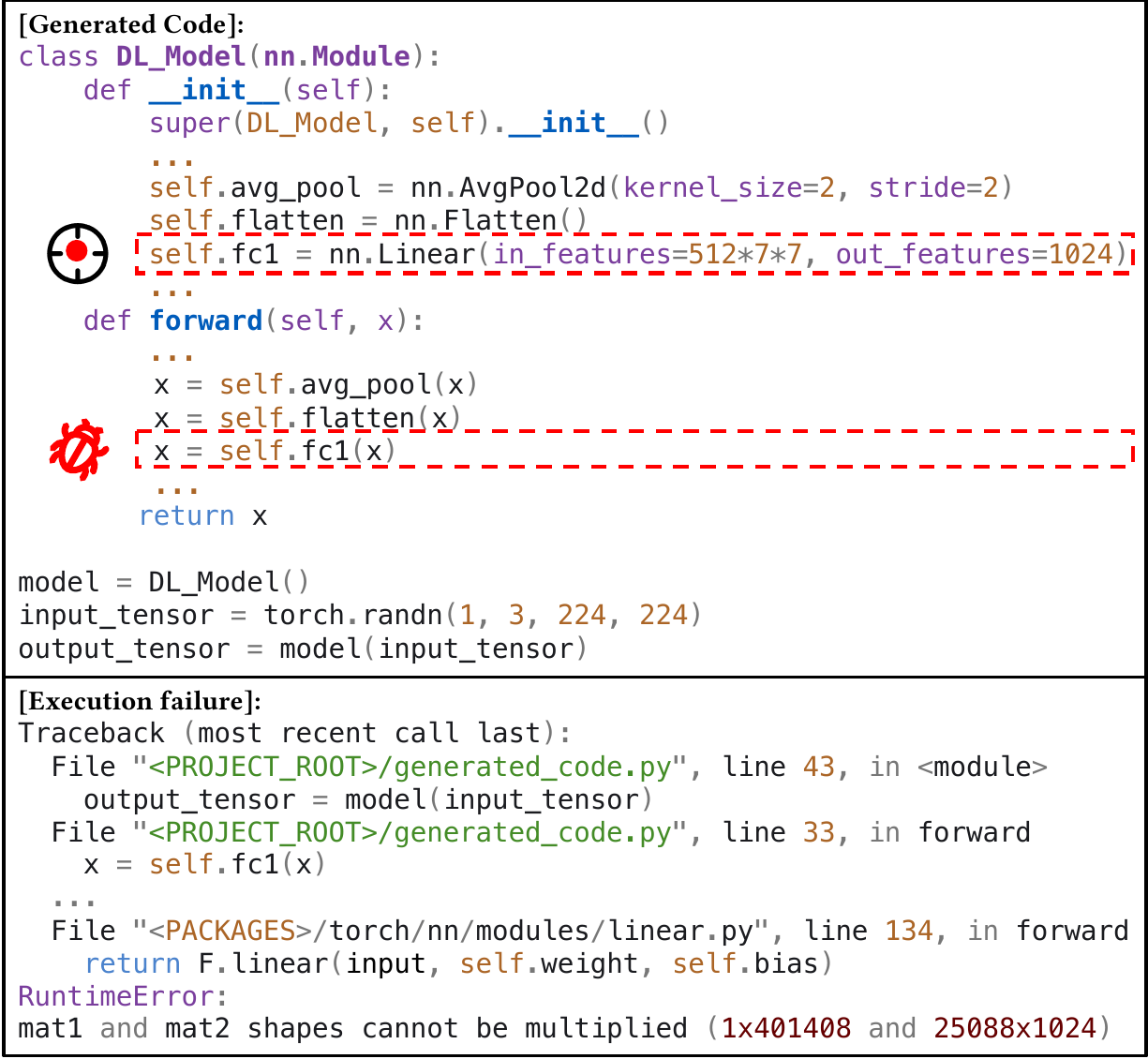}
    \caption{Example of a runtime execution failure in DL program generated by the vanilla Gemma-12B model.}
    \label{fig:executability_failure_label}
\end{figure}

%% file: figures/overview.tex
\begin{figure}[h]
    \centering
    \includegraphics[width=\linewidth]{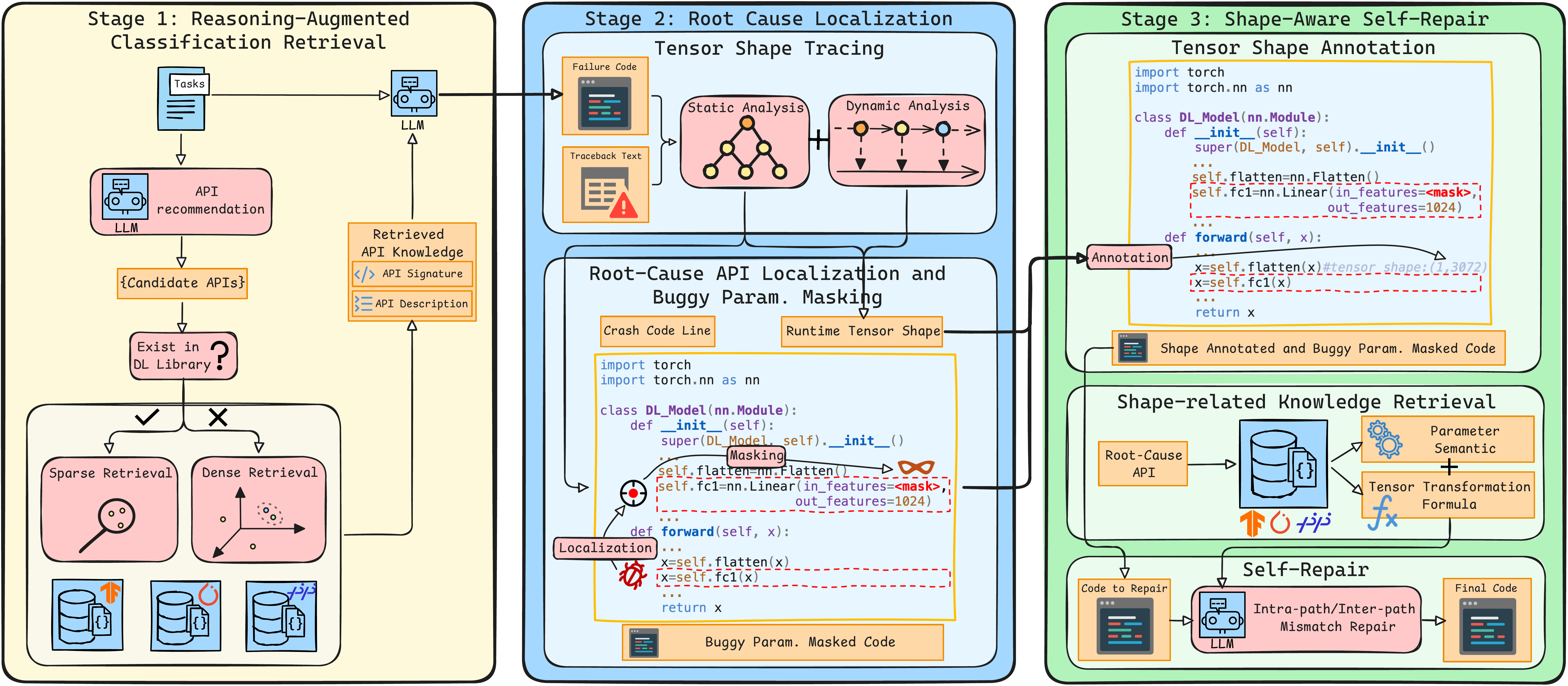}
    \caption{Overview of \textsc{DeepGroundRepair}, from API-grounded generation through root-cause localization to shape-aware self-repair.}
    \label{fig:overview}
\end{figure}

%% file: algorithm/stage1.tex
\begin{algorithm}[h]
    \caption{Reasoning-Augmented Classification Retrieval}
    \label{alg:stage1}
    
    \begin{tabular*}{\linewidth}{@{}r@{\;}l@{\extracolsep{\fill}}r@{}}
    \multicolumn{3}{@{}l}{\textbf{Input:} Task requirement $R$, DL API document database $D$}\\
    \multicolumn{3}{@{}l}{\textbf{Output:} Draft DL code $\mathit{draft\_code}$}\\
    \addlinespace[3pt]

    1 & $\mathit{candidate\_apis} = \texttt{api\_recommendation}(R)$
      & \texttt{/* API Recommendation */}\\
    2 & \textbf{for each} $\mathit{api}$ \textbf{in} $\mathit{candidate\_apis}$ \textbf{do}
      & \multirow{2}{*}{$\left.\rule{0pt}{2.8ex}\right\}$ \texttt{/* API Existence Verification */}}\\
    3 & \quad $\mathit{exists} = \texttt{api\_existence\_check}(\mathit{api})$
      & \\
    4 & \quad \textbf{if} $\mathit{exists}$ \textbf{then}
      & \multirow{4}{*}{$\left.\rule{0pt}{5.6ex}\right\}$ \texttt{/* Classification Retrieval */}}\\
    5 & \quad\quad $\mathit{docs} = \texttt{sparse\_retrieval}(\mathit{api}, D)$
      & \\
    6 & \quad \textbf{else} $\mathit{docs} = \texttt{dense\_retrieval}(\mathit{api}, D)$
      & \\
    7 & \quad $\mathit{api\_docs} = \texttt{merge}(\mathit{api\_docs}, \mathit{docs})$
      & \\
    8 & $\mathit{draft\_code} = \texttt{LLM}(R, \mathit{api\_docs})$
      & \texttt{/* Generate Draft Code */}\\
    9 & \textbf{return} $\mathit{draft\_code}$ & \\

    \end{tabular*}
\end{algorithm}

%% file: algorithm/stage2.tex
\begin{algorithm}[h]
    \caption{Root-Cause Localization}
    \label{alg:stage2}
    \begin{tabular*}{\linewidth}{@{}r@{\;}l@{\extracolsep{\fill}}r@{}}
    \multicolumn{3}{@{}l}{\textbf{Input:} Draft DL code $\mathit{draft\_code}$}\\
    \multicolumn{3}{@{}l}{\textbf{Output:} Masked DL code $\mathit{masked\_code}$, shape trace $\mathit{shape\_trace}$}\\
    \addlinespace[3pt]

    \multicolumn{3}{@{}l}{\quad\texttt{/* Tensor Shape Tracing */}}\\
    1 & $\mathit{obs\_points} = \texttt{static\_analysis}(\mathit{draft\_code})$ & \\
    2 & $(\mathit{shape\_trace}, \mathit{crash\_line}) = \texttt{dynamic\_analysis}(\mathit{draft\_code}, \mathit{obs\_points})$ & \\
    \multicolumn{3}{@{}l}{\quad\texttt{/* Root-Cause API Localization */}}\\
    3 & $\mathit{root\_api} = \texttt{localize\_root\_api}(\mathit{draft\_code}, \mathit{shape\_trace}, \mathit{crash\_line})$ & \\
    \multicolumn{3}{@{}l}{\quad\texttt{/* Buggy Parameter Masking */}}\\
    4 & $\mathit{masked\_code} = \texttt{mask\_buggy\_params}(\mathit{draft\_code}, \mathit{root\_api})$ & \\
    5 & \textbf{return} $(\mathit{masked\_code}, \mathit{shape\_trace})$ & \\
    \end{tabular*}
\end{algorithm}

%% file: algorithm/stage3.tex
\begin{algorithm}[h]
    \caption{Shape-Aware Self-Repair}
    \label{alg:stage3}
    \begingroup
    \small
    \begin{tabular*}{\linewidth}{@{}r@{\;}l@{\extracolsep{\fill}}r@{}}
    \multicolumn{3}{@{}l}{\textbf{Input:} Root-cause API $a$, Masked parameters $p$, Tensor-shape trace $T$, DL API document database $D$}\\
    \multicolumn{3}{@{}l}{\textbf{Output:} Repaired DL code $\mathit{repaired\_code}$}\\
    \addlinespace[3pt]

    1  & $T' = \texttt{tensor\_shape\_annotation}(T)$ 
       & \texttt{/* Runtime Tensor Shape Annotation */}\\
    2  & $\mathit{mismatch} = \texttt{mismatch\_analysis}(T', a, p)$ 
       & \texttt{/* Intra- vs Inter-path Detection */}\\
    3  & \textbf{if} $\mathit{mismatch} = \texttt{intra-path}$ \textbf{then}
       & \multirow{2}{*}{$\left.\rule{0pt}{4.2ex}\right\}$ \texttt{/* Intra-path Mismatch Repair */}}\\
    4  & \quad $\mathit{repaired\_code} = \texttt{LLM\_repair}(T', p)$ & \\

    5  & \textbf{else}
       & \multirow{8}{*}{$\left.\rule{0pt}{11.5ex}\right\}$ \texttt{/* Inter-path Mismatch Repair */}}\\
    6  & \quad $\mathit{shape\_kb} = \texttt{retrieve\_shape\_knowledge}(a, D)$ & \\
    7  & \quad $\mathit{constraints} = \texttt{build\_shape\_constraints}(\mathit{shape\_kb}, T')$ & \\
    8  & \quad $\Theta = \texttt{generate\_candidates\_parameters}(p, \mathit{constraints})$ & \\
    9  & \quad \textbf{for each} $\theta \in \Theta$ \textbf{do} & \\
    10 & \quad\quad $\mathit{valid} = \texttt{parameters\_solver}(\theta, \mathit{constraints}, T')$ & \\
    11 & \quad\quad \textbf{if} $\mathit{valid}$ \textbf{then} $\mathit{valid\_set}.\texttt{add}(\theta)$ & \\
    12 & \quad $\mathit{repaired\_code} = \texttt{AST\_based\_filling}(\mathit{valid\_set}, T', a)$ & \\
    13 & \textbf{return} $\mathit{repaired\_code}$ & \\
   \end{tabular*}
   \endgroup
\end{algorithm}

%% file: figures/motivation/code_executability/inter_path_mismatch.tex
\begin{figure}[h]
    \centering
    \includegraphics[width=0.78\textwidth]{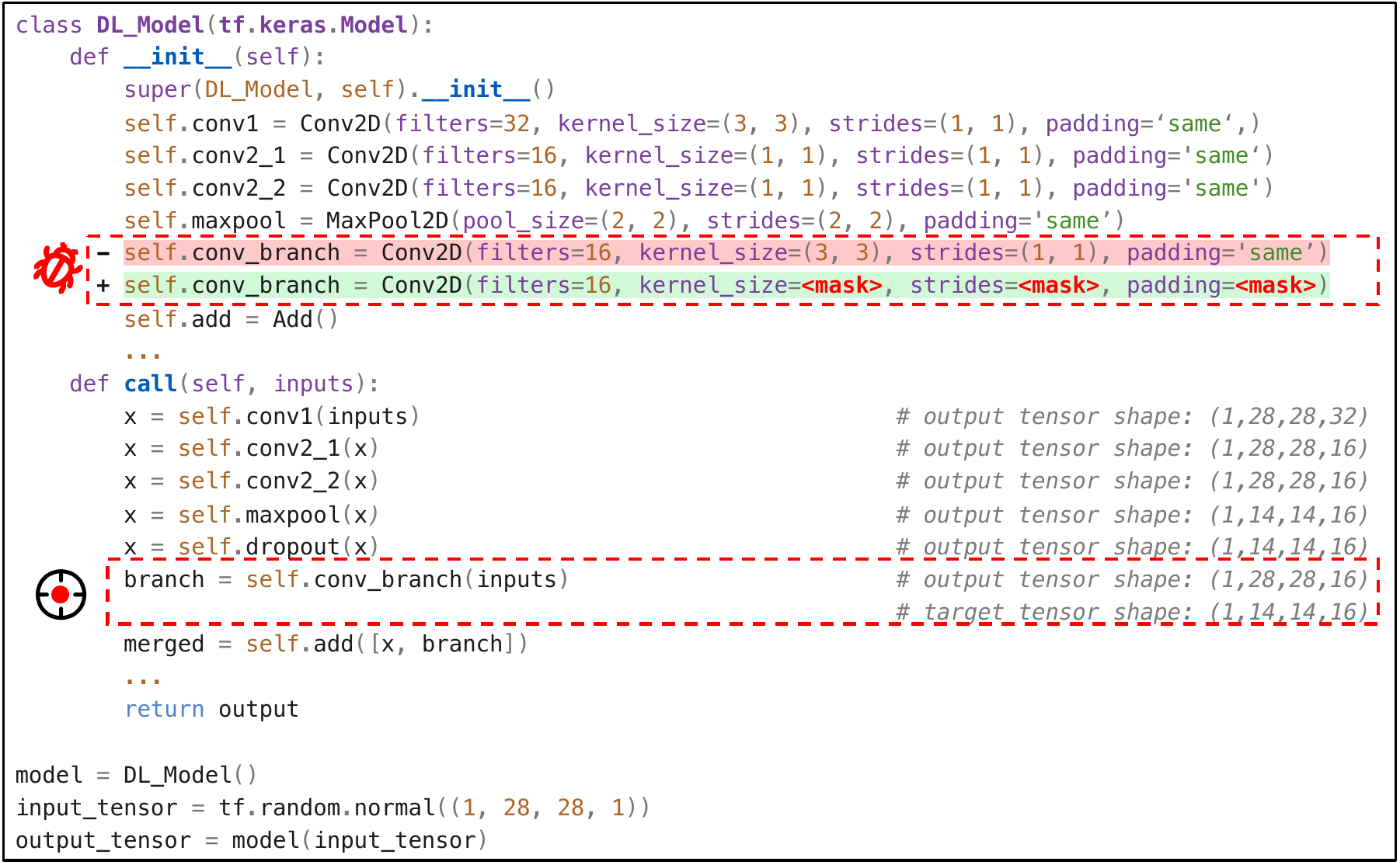}
    \caption{Example of an inter-path mismatch in a program generated by Gemma-12B. Stage~2 masks the potentially buggy parameters during root-cause API initialization in \texttt{\_\_init\_\_}, while Stage~3 annotates tensor shapes along the computation paths in \texttt{call}.}
    \label{fig:inter_path_mismatch}
\end{figure}

%% file: tables/method/shape_related_knowledge.tex
\begin{figure}[H]
    \centering
    \includegraphics[width=0.6\textwidth]{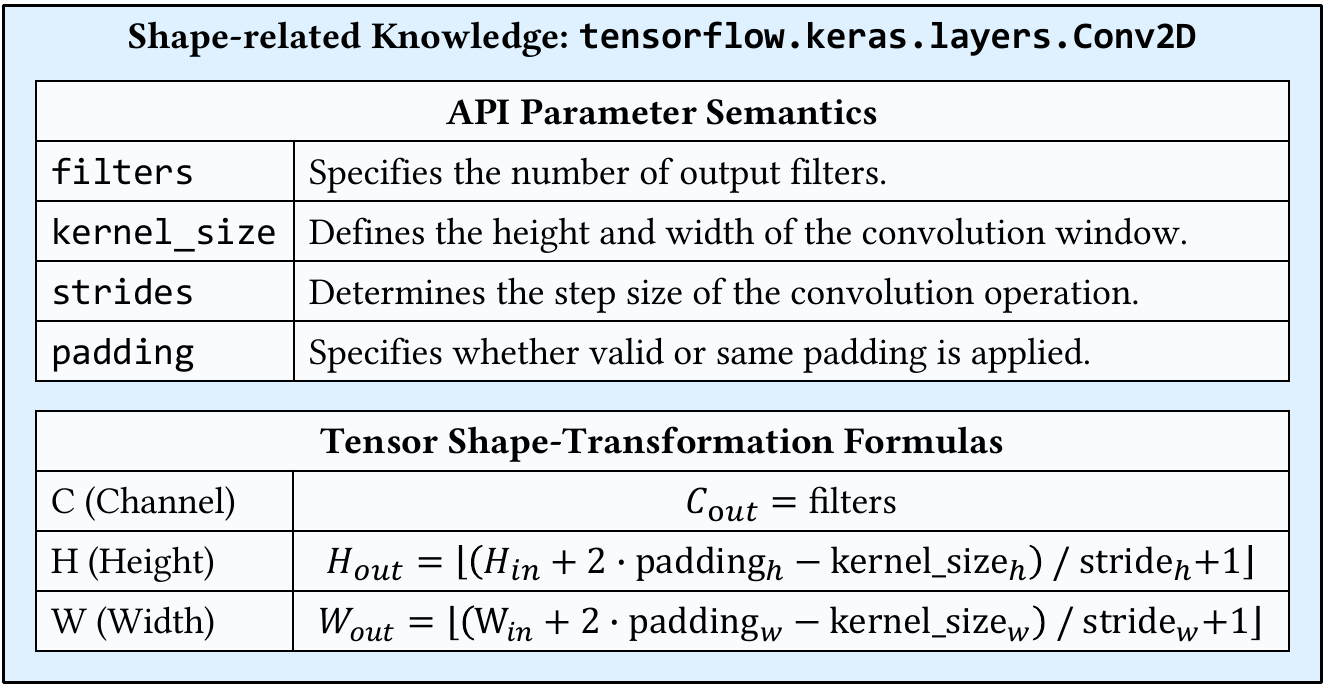}
    \caption{Example of shape-related knowledge of \texttt{tensorflow.keras.layers.Conv2D}.}
    \label{fig:shape_related_knowledge}
\end{figure}

%% file: tables/studied_llms.tex
\begin{table}[h]
  \centering
  \caption{Studied LLMs.}
  \label{tab:LLMs}
  \resizebox{1\linewidth}{!}{%
    \begin{tabular}{c|c|c|c|c|c}
    \toprule
    \textbf{Model} & \textbf{Abbr.} & \textbf{Release Time} & \textbf{Size} & \textbf{Team} & \textbf{Open Source?} \\
    \midrule

    Gemma-3-4B-it~\cite{gemma3}   & Gemma-4B   & 2025.03 & 4B  & Google       & \ding{51} \\
    Gemma-3-12B-it~\cite{gemma3}  & Gemma-12B  & 2025.03 & 12B & Google       & \ding{51} \\
    Gemma-3-27B-it~\cite{gemma3}  & Gemma-27B  & 2025.03 & 27B & Google       & \ding{51} \\

    Ministral-3-3B-Instruct~\cite{Ministral3}      & Mistral-3B & 2026.01 & 3B  & Mistral AI   & \ding{51} \\
    Ministral-3-14B-Instruct~\cite{Ministral3}      & Mistral-14B & 2026.01 & 14B  & Mistral AI   & \ding{51} \\
    Mistral-Small-3.2-24B-Instruct~\cite{mistralaimistralsmall3224Binstruct2506huggingface}     & Mistral-24B& 2025.06 & 24B & Mistral AI  & \ding{51} \\
    \bottomrule
  \end{tabular}

  }
\end{table}

%% file: tables/results/RQ1/RQ1_issues.tex
\begin{table*}[h]
  \centering
  \caption{\textit{API Hallucination Count} and \textit{API Parameter Misuse Count} per 10 program files across models and DL libraries. \textsc{DeepGR} denotes \textsc{DeepGroundRepair}, and B1--B4 denote Baselines~1--4, respectively. ``Change'' denotes the performance difference between \textsc{DeepGR} and each baseline, calculated based on the values in the ``Overall Avg.'' row. The ``\textcolor{blue}{$\downarrow$}'' and ``\textcolor{red}{$\uparrow$}'' arrows indicate fewer and more issues, respectively, and bold values indicate the best result in each row.}
  \label{tab:rq1_sc_combined}
  \setlength{\tabcolsep}{2.5pt}
  \resizebox{\linewidth}{!}{
  \begin{tabular}{
    @{}
    l
    ccccc
    @{\hspace{4pt}}
    ccccc
    @{\hspace{4pt}}
    ccccc
    @{}
  }
    \toprule
    &
    \multicolumn{5}{c}{\textbf{TensorFlow}}
    & \multicolumn{5}{c}{\textbf{PyTorch}}
    & \multicolumn{5}{c}{\textbf{PaddlePaddle}} \\
    \cmidrule(lr){2-6}
    \cmidrule(lr){7-11}
    \cmidrule(l){12-16}

    \multirow[c]{-2.2}{*}{\textbf{Model}}
    & B1 & B2 & B3 & B4 & \textsc{DeepGR}
    & B1 & B2 & B3 & B4 & \textsc{DeepGR}
    & B1 & B2 & B3 & B4 & \textsc{DeepGR} \\
    \toprule

    \rowcolor{black!8}
    \multicolumn{16}{c}{\textbf{\textit{API Hallucination Count}}} \\
    \midrule

    Gemma-4B
    & 0.00 & 3.84 & 0.00 & 2.63 & \textbf{0.00}
    & 9.50 & 4.50 & 8.30 & 4.10 & \textbf{3.74}
    & 34.8 & 22.7 & 30.9 & 21.7 & \textbf{11.7} \\

    Gemma-12B
    & 1.60 & 0.40 & 1.60 & 0.40 & \textbf{0.40}
    & 3.80 & 1.30 & 1.30 & 1.10 & 1.60
    & 3.90 & 8.70 & 4.30 & 6.70 & 4.29 \\

    Gemma-27B
    & 0.60 & 0.50 & 0.40 & 0.50 & \textbf{0.00}
    & 3.70 & 1.40 & 0.30 & 0.30 & 0.80
    & 9.80 & 8.40 & 6.70 & 6.00 & \textbf{1.40} \\
    
    \rowcolor{cyan!12}
    \textit{Average}
    & 0.73 & 1.58 & 0.67 & 1.18 & \textbf{0.13}
    & 5.67 & 2.40 & 3.30 & 1.83 & 2.05
    & 16.2 & 13.3 & 14.0 & 11.5 & \textbf{5.80} \\
    
    \midrule
    \addlinespace[5pt]

    Mistral-3B
    & 0.41 & 0.00 & 0.00 & 0.00 & \textbf{0.00}
    & 0.90 & 0.80 & 0.80 & 0.80 & 0.84
    & 13.1 & 8.60 & 8.60 & 8.60 & \textbf{2.47} \\

    Mistral-14B
    & 0.20 & 0.10 & 0.10 & 0.20 & \textbf{0.00}
    & 0.20 & 0.00 & 0.00 & 0.00 & 0.30
    & 0.60 & 0.40 & 0.40 & 0.40 & \textbf{0.00} \\

    Mistral-24B
    & 0.00 & 0.00 & 0.00 & 0.00 & 0.20
    & 0.31 & 0.00 & 0.00 & 0.00 & 0.10
    & 0.40 & 1.10 & 1.10 & 1.10 & \textbf{0.10} \\
    
    \rowcolor{cyan!12}
    \textit{Average}
    & 0.20 & 0.03 & 0.03 & 0.07 & 0.07
    & 0.47 & 0.27 & 0.27 & 0.27 & 0.41
    & 4.70 & 3.37 & 3.37 & 3.37 & \textbf{0.86} \\

    \midrule
    \addlinespace[5pt]

    \rowcolor{cyan!12}
    \textbf{\textit{Overall Avg.}}
    & 0.47 & 0.81 & 0.35 & 0.62 & \textbf{0.10}
    & 3.07 & 1.33 & 1.78 & 1.05 & 1.23
    & 10.43 & 8.32 & 8.67 & 7.42 & \textbf{3.33} \\
    \rowcolor{cyan!12}
    \textbf{\textit{Change}}
    & \textcolor{blue}{$\downarrow$\textbf{0.37}} & \textcolor{blue}{$\downarrow$\textbf{0.71}} & \textcolor{blue}{$\downarrow$\textbf{0.25}} & \textcolor{blue}{$\downarrow$\textbf{0.52}} & --
    & \textcolor{blue}{$\downarrow$\textbf{1.84}} & \textcolor{blue}{$\downarrow$\textbf{0.10}} & \textcolor{blue}{$\downarrow$\textbf{0.55}} & \textcolor{red}{$\uparrow$0.18} & --
    & \textcolor{blue}{$\downarrow$\textbf{7.10}} & \textcolor{blue}{$\downarrow$\textbf{4.99}} & \textcolor{blue}{$\downarrow$\textbf{5.34}} & \textcolor{blue}{$\downarrow$\textbf{4.09}} & -- \\

    \cmidrule{1-16}

    \rowcolor{black!8}
    \multicolumn{16}{c}{\textbf{\textit{API Parameter Misuse Count}}} \\
    \midrule

    Gemma-4B
    & 3.50 & 0.91 & 3.90 & 0.91 & \textbf{0.41}
    & 0.00 & 0.00 & 0.00 & 0.00 & \textbf{0.00}
    & 76.8 & 3.94 & 65.6 & 4.04 & 11.2 \\

    Gemma-12B
    & 3.80 & 3.20 & 4.00 & 3.20 & \textbf{1.52}
    & 0.00 & 0.00 & 0.30 & 0.00 & 0.30
    & 14.3 & 4.00 & 10.2 & 2.10 & 2.86 \\

    Gemma-27B
    & 6.80 & 1.00 & 5.60 & 1.20 & \textbf{0.20}
    & 0.00 & 0.00 & 0.00 & 0.00 & 0.10
    & 0.10 & 2.00 & 0.00 & 1.90 & 0.70 \\
    
    \rowcolor{cyan!12}
    \textit{Average}
    & 4.70 & 1.70 & 4.50 & 1.77 & \textbf{0.71}
    & 0.00 & 0.00 & 0.10 & 0.00 & 0.13
    & 30.4 & 3.31 & 25.27 & 2.68 & 4.92 \\
    

    \midrule
    \addlinespace[5pt]

    Mistral-3B
    & 5.98 & 4.60 & 4.69 & 4.60 & \textbf{1.58}
    & 0.00 & 0.00 & 0.00 & 0.00 & \textbf{0.00}
    & 9.20 & 5.80 & 5.80 & 5.80 & \textbf{1.13} \\

    Mistral-14B
    & 5.00 & 5.05 & 5.00 & 5.05 & \textbf{0.71}
    & 0.20 & 0.20 & 0.20 & 0.20 & \textbf{0.00}
    & 0.80 & 0.80 & 0.80 & 0.80 & \textbf{0.41} \\

    Mistral-24B
    & 7.00 & 7.00 & 7.00 & 7.00 & \textbf{0.61}
    & 0.20 & 0.00 & 0.00 & 0.00 & \textbf{0.00}
    & 0.30 & 0.10 & 0.10 & 0.10 & 0.20 \\
    
    \rowcolor{cyan!12}
    \textit{Average}
    & 5.99 & 5.55 & 5.56 & 5.55 & \textbf{0.97}
    & 0.13 & 0.07 & 0.07 & 0.07 & \textbf{0.00}
    & 3.43 & 2.23 & 2.23 & 2.23 & \textbf{0.58} \\
    

    \midrule
    \addlinespace[5pt]
 
    \rowcolor{cyan!12} 
    \textbf{\textit{Overall Avg.}}
    & 5.35 & 3.63 & 5.03 & 3.66 & \textbf{0.84}
    & 0.07 & 0.03 & 0.08 & 0.03 & 0.07
    & 16.92 & 2.77 & 13.75 & 2.46 & 2.75 \\
    
    \rowcolor{cyan!12}
    \textbf{\textit{Change}}
    & \textcolor{blue}{$\downarrow$\textbf{4.51}} & \textcolor{blue}{$\downarrow$\textbf{2.79}} & \textcolor{blue}{$\downarrow$\textbf{4.19}} & \textcolor{blue}{$\downarrow$\textbf{2.82}} & --
    & 0.00 & \textcolor{red}{$\uparrow$0.04} & \textcolor{blue}{$\downarrow$\textbf{0.01}} & \textcolor{red}{$\uparrow$0.04} & --
    & \textcolor{blue}{$\downarrow$\textbf{14.17}} & \textcolor{blue}{$\downarrow$\textbf{0.02}} & \textcolor{blue}{$\downarrow$\textbf{11.00}} & \textcolor{red}{$\uparrow$0.29} & -- \\
    \bottomrule
  \end{tabular}
  }
\end{table*}

%% file: tables/results/RQ1/RQ1CER.tex

\begin{table}[h]
  \caption{\textit{Code Executability Rate} (\textit{CER}, \%) of each model across DL libraries and methods. \textsc{DeepGR} denotes \textsc{DeepGroundRepair}, and B1--B4 denote Baselines~1--4, respectively. ``$\Delta$ (\%)'' reports the relative change of \textsc{DeepGR} over each corresponding baseline using the values in the ``Overall Avg.'' row. The ``\textcolor{blue}{$\uparrow$}'' arrow indicates improvements, and bold values indicate the best result in each row.}
  \label{tab:rq1_cer}
  \centering
  \setlength{\tabcolsep}{3.2pt}
  \resizebox{1\linewidth}{!}{%
  \begin{tabular}{
    @{\extracolsep{\fill}}
    l
    ccccc
    @{\hspace{8pt}}
    ccccc
    @{\hspace{8pt}}
    ccccc
    @{}
  }
    \toprule
    \multirow{2}[2]{*}{\textbf{Model}}  
    & \multicolumn{5}{c}{\textbf{TensorFlow}}
    & \multicolumn{5}{c}{\textbf{PyTorch}}
    & \multicolumn{5}{c}{\textbf{PaddlePaddle}} \\
    \cmidrule(lr){2-6}
    \cmidrule(lr){7-11}
    \cmidrule(l){12-16}
    & B1 & B2 & B3 & B4 & \textsc{DeepGR}
    & B1 & B2 & B3 & B4 & \textsc{DeepGR}
    & B1 & B2 & B3 & B4 & \textsc{DeepGR} \\
    \midrule
    Gemma-4B          & 36 & 20 & 38 & 20 & \textbf{46} &  6 &  9 &  6 &  9 & \textbf{25} &  0 &  0 &  0 &  6 & \textbf{12} \\
    Gemma-12B         & 36 & 26 & 43 & 32 & \textbf{59} & 18 & 19 & 21 & 24 & \textbf{34} &  2 &  2 &  5 &  5 & \textbf{21} \\
    Gemma-27B         & 44 & 38 & 55 & 49 & \textbf{74} & 19 & 28 & 28 & 34 & \textbf{39} & 11 &  5 & 21 & 12 & \textbf{36} \\

    \rowcolor{cyan!12}
    \textit{Average}
    & 38.7 & 28.0 & 45.3 & 33.7 & \textbf{59.7}
    & 14.3 & 18.7 & 18.3 & 22.3 & \textbf{32.7}
    &  4.3 &  2.3 &  8.7 &  7.7 & \textbf{23.0} \\
    \midrule
    \addlinespace[5pt]
    Mistral-3B      & 13 & 29 & 33 & 44 & \textbf{45} &  3 & 15 &  8 & 21 & \textbf{38} &  1 &  4 &  6 & 12 & \textbf{20} \\
    Mistral-14B     & 51 & 47 & 63 & 66 & 59 & 31 & 32 & 49 & 43 & \textbf{54} & 27 & 14 & 35 & 29 & \textbf{48} \\
    Mistral-24B & 47 & 60 & 75 & 83 & 66 & 25 & 33 & 47 & 57 & \textbf{57} & 32 & 25 & 50 & 43 & \textbf{52} \\

    \rowcolor{cyan!12}
    \textit{Average}
    & 37.0 & 45.3 & \textbf{57.0} & 64.3 & 56.7
    & 19.7 & 26.7 & 34.7 & 40.3 & \textbf{49.7}
    & 20.0 & 14.3 & 30.3 & 28.0 & \textbf{40.0} \\
    \rowcolor{cyan!12}

    \midrule
    \addlinespace[5pt]
    \rowcolor{cyan!12}
    \textbf{\textit{Overall Avg.}}
    & 37.8 & 36.7 & 51.2 & 49.0 & \textbf{58.2}
    & 17.0 & 22.7 & 26.5 & 31.3 & \textbf{41.2}
    & 12.2 &  8.3 & 19.5 & 17.8 & \textbf{31.5} \\
    \rowcolor{cyan!12}
    \textbf{$\Delta$ (\%)}
    & \textcolor{blue}{$\uparrow$\textbf{54.0}} & \textcolor{blue}{$\uparrow$\textbf{58.6}} & \textcolor{blue}{$\uparrow$\textbf{13.7}} & \textcolor{blue}{$\uparrow$\textbf{18.8}} & --
    & \textcolor{blue}{$\uparrow$\textbf{142.4}} & \textcolor{blue}{$\uparrow$\textbf{81.5}} & \textcolor{blue}{$\uparrow$\textbf{55.5}} & \textcolor{blue}{$\uparrow$\textbf{31.6}} & --
    & \textcolor{blue}{$\uparrow$\textbf{155.7}} & \textcolor{blue}{$\uparrow$\textbf{275.9}} & \textcolor{blue}{$\uparrow$\textbf{60.0}} & \textcolor{blue}{$\uparrow$\textbf{75.3}} & -- \\
    \bottomrule
  \end{tabular}
  }
\end{table}

%% file: tables/results/RQ1/rq1_semantic_consistency_rate.tex
\begin{table}[h]
  \caption{\textit{Semantic Consistency Rate} (\textit{SCR},\%) of each model across DL libraries and methods. \textsc{DeepGR} denotes \textsc{DeepGroundRepair}, and B3--B4 denote Baselines~3--4, respectively. ``$\Delta$ (\%)'' reports the relative change of \textsc{DeepGR} over each corresponding baseline using the values in the ``Overall Avg.'' row. The ``\textcolor{blue}{$\uparrow$}'' arrow indicate improvements, and bold values indicate the best result in each row.}
  \label{tab:rq1_semantic_consistency_rate}
  \centering
  \begin{tabular}{
    l
    ccc
    @{\hspace{10pt}}
    ccc
    @{\hspace{10pt}}
    ccc
  }
    \toprule
    \multirow{2}[2]{*}{\textbf{Model}}  
    & \multicolumn{3}{c}{\textbf{TensorFlow}}
    & \multicolumn{3}{c}{\textbf{PyTorch}}
    & \multicolumn{3}{c}{\textbf{PaddlePaddle}} \\
    \cmidrule(lr){2-4}
    \cmidrule(lr){5-7}
    \cmidrule(l){8-10}
    & B3 & B4 & \textsc{DeepGR}
    & B3 & B4 & \textsc{DeepGR}
    & B3 & B4 & \textsc{DeepGR} \\
    \midrule
    Gemma-4B  & \textbf{97.0} & 89.9 & 95.9 & 76.0 & 79.0 & \textbf{94.9} & 70.0 & 91.9 & \textbf{96.0} \\
    Gemma-12B & 86.0 & 89.0 & \textbf{92.9} & 91.0 & 90.0 & \textbf{99.0} & 71.0 & \textbf{93.0} & 87.8 \\
    Gemma-27B & 86.0 & 91.0 & \textbf{97.0} & 86.0 & 91.0 & \textbf{97.0} & 83.0 & 89.0 & \textbf{100.0} \\
    
    \rowcolor{cyan!12}
    \textit{Average}
    & 89.67 & 89.97 & \textbf{95.27}
    & 84.33 & 86.67 & \textbf{96.97}
    & 74.67 & 91.30 & \textbf{94.60} \\
    
    \midrule
    \addlinespace[5pt]
    Mistral-3B & 61.9 & 61.0 & \textbf{88.4} & 72.7 & 56.0 & \textbf{96.8} & 73.0 & 47.0 & \textbf{95.8} \\
    Mistral-14B & 74.0 & 75.8 & \textbf{98.0} & 83.0 & 87.0 & \textbf{94.0} & 81.0 & 79.0 & \textbf{91.8} \\
    Mistral-24B & 76.0 & 83.0 & \textbf{91.9} & 79.6 & 88.0 & \textbf{97.0} & 78.0 & 83.0 & \textbf{91.0} \\
    
    \rowcolor{cyan!12}
    \textit{Average}
    & 70.63 & 73.27 & \textbf{92.77}
    & 78.43 & 77.00 & \textbf{95.93}
    & 77.33 & 69.67 & \textbf{92.87} \\
    
    \midrule
    \addlinespace[5pt]
    \rowcolor{cyan!12}
    \textbf{\textit{Overall Avg.}}
    & 80.15 & 81.62 & \textbf{94.02}
    & 81.38 & 81.83 & \textbf{96.45}
    & 76.00 & 80.48 & \textbf{93.73} \\
    \rowcolor{cyan!12}
    \textbf{\textit{$\Delta$} (\%)}
    & \textcolor{blue}{$\uparrow$\textbf{17.3}} & \textcolor{blue}{$\uparrow$\textbf{15.2}} & --
    & \textcolor{blue}{$\uparrow$\textbf{18.5}} & \textcolor{blue}{$\uparrow$\textbf{17.9}} & --
    & \textcolor{blue}{$\uparrow$\textbf{23.3}} & \textcolor{blue}{$\uparrow$\textbf{16.5}} & -- \\
    
    \bottomrule
  \end{tabular}
\end{table}

%% file: tables/results/RQ2/rq2_issues.tex
\begin{table*}[h]
  \centering
  \caption{\textit{API Hallucination Count} and \textit{API Parameter Misuse Count} per 10 program files for the complete \textsc{DeepGroundRepair} pipeline and its three ablated variants, denoted in the table by \textsc{DeepGR} and V1--V3, respectively. ``\textit{Average}'' denotes the arithmetic mean across the two models. Superscript values denote the performance difference between each variant and \textsc{DeepGR}, calculated from the ``\textit{Average}'' values. The ``\textcolor{blue}{$\downarrow$}'' and ``\textcolor{red}{$\uparrow$}'' symbols indicate lower and higher error counts, respectively.}

  \label{tab:rq2_sc_combined}
  \setlength{\tabcolsep}{2.5pt}
  \resizebox{\linewidth}{!}{
    \begin{tabular}{
      @{}
      l
      cccc
      @{\hspace{4pt}}
      cccc
      @{\hspace{4pt}}
      cccc
      @{}
    }
      \toprule
      & \multicolumn{4}{c}{\textbf{TensorFlow}}
      & \multicolumn{4}{c}{\textbf{PyTorch}}
      & \multicolumn{4}{c}{\textbf{PaddlePaddle}} \\
      \cmidrule(lr){2-5}
      \cmidrule(lr){6-9}
      \cmidrule(l){10-13}

      \multirow[c]{-2.2}{*}{\textbf{Model}}
      & \textsc{DeepGR} & V1 & V2 & V3
      & \textsc{DeepGR} & V1 & V2 & V3
      & \textsc{DeepGR} & V1 & V2 & V3 \\
      \midrule

      \rowcolor{black!8}
      \multicolumn{13}{c}{\textbf{\textit{API Hallucination Count}}} \\
      \midrule

      Gemma-12B
      & 0.40 & 1.60 & 0.51 & 0.20
      & 1.60 & 2.20 & 1.50 & 1.90
      & 4.29 & 4.30 & 4.00 & 4.90 \\

      Mistral-14B
      & 0.00 & 0.00 & 0.00 & 0.00
      & 0.30 & 0.30 & 0.60 & 0.50
      & 0.00 & 2.60 & 0.10 & 0.10 \\

      \midrule
      \rowcolor{cyan!12}
      \textit{\textbf{Average}}
      & 0.20
      & 0.80\textsuperscript{\textcolor{red}{$\uparrow$0.60}}
      & 0.25\textsuperscript{\textcolor{red}{$\uparrow$0.05}}
      & 0.10\textsuperscript{\textcolor{blue}{$\downarrow$0.10}}
      & 0.95
      & 1.25\textsuperscript{\textcolor{red}{$\uparrow$0.30}}
      & 1.05\textsuperscript{\textcolor{red}{$\uparrow$0.10}}
      & 1.20\textsuperscript{\textcolor{red}{$\uparrow$0.25}}
      & 2.14
      & 3.45\textsuperscript{\textcolor{red}{$\uparrow$1.31}}
      & 2.05\textsuperscript{\textcolor{blue}{$\downarrow$0.09}}
      & 2.50\textsuperscript{\textcolor{red}{$\uparrow$0.36}} \\

      \midrule

      \rowcolor{black!8}
      \multicolumn{13}{c}{\textbf{\textit{API Parameter Misuse Count}}} \\
      \midrule

      Gemma-12B
      & 1.51 & 3.80 & 1.41 & 1.82
      & 0.30 & 0.30 & 0.30 & 0.00
      & 2.86 & 12.40 & 2.30 & 3.27 \\

      Mistral-14B
      & 0.71 & 1.00 & 0.60 & 0.61
      & 0.00 & 0.10 & 0.40 & 0.20
      & 0.41 & 2.70 & 0.00 & 0.41 \\

      \midrule
      \rowcolor{cyan!12}
      \textit{\textbf{Average}}
      & 1.11
      & 2.40\textsuperscript{\textcolor{red}{$\uparrow$1.29}}
      & 1.01\textsuperscript{\textcolor{blue}{$\downarrow$0.10}}
      & 1.21\textsuperscript{\textcolor{red}{$\uparrow$0.10}}
      & 0.15
      & 0.20\textsuperscript{\textcolor{red}{$\uparrow$0.05}}
      & 0.35\textsuperscript{\textcolor{red}{$\uparrow$0.20}}
      & 0.10\textsuperscript{\textcolor{blue}{$\downarrow$0.05}}
      & 1.63
      & 7.55\textsuperscript{\textcolor{red}{$\uparrow$5.92}}
      & 1.15\textsuperscript{\textcolor{blue}{$\downarrow$0.48}}
      & 1.84\textsuperscript{\textcolor{red}{$\uparrow$0.21}} \\

      \bottomrule
    \end{tabular}
  }
\end{table*}

%% file: tables/results/RQ2/RQ2_cer.tex
\begin{table}[h]
  \centering
  \caption{\textit{Code Executability Rate} (\textit{CER}, \%) for the complete \textsc{DeepGroundRepair} pipeline and its three variants, denoted in the table by \textsc{DeepGR} and V1--V3, respectively. ``\textit{Average}'' denotes the arithmetic mean across the two models. Superscript values denote the relative change of each variant compared with \textsc{DeepGR}. The ``\textcolor{blue}{$\uparrow$}'' and ``\textcolor{red}{$\downarrow$}'' indicate higher and lower \textit{CER} (\%), respectively.}
  \label{tab:rq2_ab}
  \small
  \setlength{\tabcolsep}{2.5pt}
  \resizebox{\linewidth}{!}{%
  \begin{tabular}{
    @{}
    l
    cccc
    @{\hspace{8pt}}
    cccc
    @{\hspace{8pt}}
    cccc
    @{}
  }
  \toprule
  & \multicolumn{4}{c}{\textbf{TensorFlow}}
  & \multicolumn{4}{c}{\textbf{PyTorch}}
  & \multicolumn{4}{c}{\textbf{PaddlePaddle}} \\
  \cmidrule(lr){2-5}
  \cmidrule(lr){6-9}
  \cmidrule(l){10-13}

  \multirow[c]{-2.2}{*}{\textbf{Model}}
  & \textsc{DeepGR} & \textsc{V1} & \textsc{V2} & \textsc{V3}
  & \textsc{DeepGR} & \textsc{V1} & \textsc{V2} & \textsc{V3}
  & \textsc{DeepGR} & \textsc{V1} & \textsc{V2} & \textsc{V3} \\
  \midrule

  Gemma-12B
  & 59 & 52 & 57 & 44
  & 34 & 38 & 21 & 17
  & 21 & 5 & 13 & 12 \\

  Mistral-14B
  & 59 & 55 & 62 & 49
  & 54 & 54 & 49 & 36
  & 48 & 31 & 43 & 31 \\

  \midrule
  \rowcolor{cyan!12}
  \textbf{\textit{Average}}
  & 59.0
  & 53.5\textsuperscript{\textcolor{red}{$\downarrow$9.3\%}}
  & 59.5\textsuperscript{\textcolor{blue}{$\uparrow$0.8\%}}
  & 46.5\textsuperscript{\textcolor{red}{$\downarrow$21.2\%}}
  & 44.0
  & 46.0\textsuperscript{\textcolor{blue}{$\uparrow$4.5\%}}
  & 35.0\textsuperscript{\textcolor{red}{$\downarrow$20.5\%}}
  & 26.5\textsuperscript{\textcolor{red}{$\downarrow$39.8\%}}
  & 34.5
  & 18.0\textsuperscript{\textcolor{red}{$\downarrow$47.8\%}}
  & 28.0\textsuperscript{\textcolor{red}{$\downarrow$18.8\%}}
  & 21.5\textsuperscript{\textcolor{red}{$\downarrow$37.7\%}} \\

  \bottomrule
  \end{tabular}
  }
\end{table}

%% file: tables/results/RQ2/rq2_semantic_consistency_rate.tex
\begin{table}[h]
  \caption{\textit{Semantic Consistency Rate} (\textit{SCR}, \%) for the complete \textsc{DeepGroundRepair} pipeline and its three variants, denoted in the table by \textsc{DeepGR} and V1--V3, respectively. ``\textit{Average}'' denotes the arithmetic mean across the two models. Superscript values denote the relative change of each variant compared with \textsc{DeepGR}. The ``\textcolor{blue}{$\uparrow$}'' and ``\textcolor{red}{$\downarrow$}'' indicate higher and lower \textit{SCR} (\%), respectively.}
  \label{tab:rq2_semantic_consistency_rate}
  \centering
  \setlength{\tabcolsep}{3.2pt}
  \resizebox{\linewidth}{!}{
  \begin{tabular}{
    @{\extracolsep{\fill}}
    l
    cccc
    @{\hspace{8pt}}
    cccc
    @{\hspace{8pt}}
    cccc
    @{}
  }
    \toprule
    & \multicolumn{4}{c}{\textbf{TensorFlow}}
    & \multicolumn{4}{c}{\textbf{PyTorch}}
    & \multicolumn{4}{c}{\textbf{PaddlePaddle}} \\
    \cmidrule(lr){2-5}
    \cmidrule(lr){6-9}
    \cmidrule(l){10-13}
    \multirow[c]{-2.2}{*}{\textbf{Model}}
    & \textsc{DeepGR} & \textsc{V1} & \textsc{V2} & \textsc{V3}
    & \textsc{DeepGR} & \textsc{V1} & \textsc{V2} & \textsc{V3}
    & \textsc{DeepGR} & \textsc{V1} & \textsc{V2} & \textsc{V3} \\
    \midrule
    Gemma-12B & 92.9 & 90.0 & 75.8 & 97.0 & 99.0 & 95.0 & 84.0 & 94.0 & 87.8 & 78.0 & 81.0 & 84.7 \\
    Mistral-14B & 98.0 & 92.0 & 74.0 & 94.9 & 94.0 & 94.0 & 83.0 & 92.0 & 91.8 & 86.0 & 83.0 & 87.7 \\

    \midrule
    \rowcolor{cyan!12}
    \textbf{\textit{Average}}
    & 95.5
    & 91.0\textsuperscript{\textcolor{red}{$\downarrow$4.7\%}}
    & 74.9\textsuperscript{\textcolor{red}{$\downarrow$21.6\%}}
    & 96.0\textsuperscript{\textcolor{blue}{$\uparrow$0.5\%}}
    & 96.5
    & 94.5\textsuperscript{\textcolor{red}{$\downarrow$2.1\%}}
    & 83.5\textsuperscript{\textcolor{red}{$\downarrow$13.5\%}}
    & 93.0\textsuperscript{\textcolor{red}{$\downarrow$3.6\%}}
    & 89.8
    & 82.0\textsuperscript{\textcolor{red}{$\downarrow$8.7\%}}
    & 82.0\textsuperscript{\textcolor{red}{$\downarrow$8.7\%}}
    & 86.2\textsuperscript{\textcolor{red}{$\downarrow$4.0\%}} \\
    \bottomrule
  \end{tabular}
  }
\end{table}

%% file: tables/results/RQ1/RQ1_RA.tex
\begin{table}[h]
  \caption{\textit{Precision@10} (\%) and \textit{Recall@10} (\%) for Hybrid-RAG in Baseline~2 and \textit{Reasoning-Augmented Classification Retrieval} of \textsc{DeepGroundRepair} across models and DL libraries. Here, Baseline~2 and \textsc{DeepGroundRepair} are denoted by B2 and \textsc{DeepGR}.}
  \label{tab:rq3_ra}
  \centering
  \resizebox{\textwidth}{!}{
  \begin{tabular}{
    @{\extracolsep{\fill}}
    cc
    cc
    @{\hspace{8pt}}
    cc
    @{\hspace{8pt}}
    cc
    @{}
  }
  \toprule
  \multicolumn{1}{c}{
    \multirow{2}*[-1ex]{
      \shortstack[c]{\textbf{Retrieval} \\ \textbf{Method}}
    }
  }
  &
  \multicolumn{1}{c}{
    \multirow{2}*[-1ex]{
      \shortstack[c]{\textbf{Recommender} \\ \textbf{LLM}}
    }
  }
  & \multicolumn{2}{c}{\textbf{TensorFlow}}
  & \multicolumn{2}{c}{\textbf{PyTorch}}
  & \multicolumn{2}{c}{\textbf{PaddlePaddle}} \\
  \cmidrule(lr){3-4}
  \cmidrule(lr){5-6}
  \cmidrule(l){7-8}
  & &
  \shortstack[c]{\textit{Precision@10}}
  & \shortstack[c]{\textit{Recall@10}}
  & \shortstack[c]{\textit{Precision@10}}
  & \shortstack[c]{\textit{Recall@10}}
  & \shortstack[c]{\textit{Precision@10}}
  & \shortstack[c]{\textit{Recall@10}} \\
  \midrule 
  \shortstack[c]{Hybrid-RAG of B2} 
  & --
  & 2.4
  & 3.3
  & 16.2
  & 20.9
  & 0.5
  & 1.0 \\
    \midrule
    \addlinespace[5pt]
    \multirow{6}{*}{\shortstack[c]{Stage 1 of \textsc{DeepGR}}}
      & Gemma-4B
      & 62.0 & 84.8
      & 59.7 & 77.7
      & 42.8 & 62.1 \\
    & Gemma-12B
      & 80.2 & 85.8
      & 83.0 & 79.2
      & 74.4 & 79.3 \\
    & Gemma-27B
      & 89.9 & 89.5
      & 79.9 & 79.2
      & 73.8 & 77.5 \\  
    & Mistral-3B
      & 72.9 & 85.7
      & 58.7 & 72.3
      & 50.3 & 72.2 \\
    & Mistral-14B
      & 73.6 & 94.4
      & 65.6 & 79.9
      & 59.5 & 73.8 \\
    & Mistral-24B
      & 83.7 & 80.7
      & 82.8 & 74.1
      & 81.2 & 74.8 \\
    \bottomrule
  \end{tabular}
  }
\end{table}

%% file: tables/method/stage1_example.tex
\begin{table}[h]
    \centering
    \caption{An illustrative example of \textit{Reasoning-Augmented Classification Retrieval} in PyTorch, showing how LLM-recommended APIs are verified and routed to sparse or dense retrieval based on their existence.}
    \label{tab:racr_example}
    \setlength{\tabcolsep}{4pt}
    \begin{tabular}{@{}lccl@{}}
      \toprule
      \textbf{LLM-Recommended API}
      & \textbf{Existence}
      & \textbf{Retrieval Strategy}
      & \textbf{Retrieved API} \\
      \midrule
      \texttt{torch.nn.Conv2d}
      & \ding{51}
      & sparse retrieval
      & \texttt{torch.nn.Conv2d} \\
  
      \texttt{torch.nn.MaxPool2d}
      & \ding{51}
      & sparse retrieval
      & \texttt{torch.nn.MaxPool2d} \\
  
      \texttt{torch.nn.functional.add}
      & \ding{56}
      & dense retrieval
      & \texttt{torch.add} \\
  
      \texttt{torch.nn.Linear}
      & \ding{51}
      & sparse retrieval
      & \texttt{torch.nn.Linear} \\
      \bottomrule
    \end{tabular}
\end{table}

%% file: tables/results/RQ3/rq3_api_acc.tex
\begin{table}[h]
    \centering
    \caption{Evaluation of Stage~2 using \textit{Failure-Site--Root-Cause Discrepancy Rate} (\textit{FRDR}, \%) and \textit{Root-Cause API Localization Accuracy} (\textit{RCLA}, \%) on the sampled cases. For each library, \textit{Overall} is calculated over all cases from both models.}
    \label{tab:rq3_api_localization}
    \begin{tabular}{
      @{\extracolsep{\fill}}
      l
      cc
      @{\hspace{8pt}}
      cc
      @{\hspace{8pt}}
      cc
      @{}
    }
      \toprule
      & \multicolumn{2}{c}{\textbf{TensorFlow}}
      & \multicolumn{2}{c}{\textbf{PyTorch}}
      & \multicolumn{2}{c}{\textbf{PaddlePaddle}} \\
      \cmidrule(lr){2-3}
      \cmidrule(lr){4-5}
      \cmidrule(l){6-7}
      \multirow[c]{-2.2}{*}{\textbf{Model}}
      & \textit{FRDR}
      & \textit{RCLA}
      & \textit{FRDR}
      & \textit{RCLA}
      & \textit{FRDR}
      & \textit{RCLA} \\
      \midrule
      Gemma-12B & 100.0 & 80.0 & 92.0 & 92.0 & 81.0 & 85.7 \\
      Mistral-14B & 100.0 & 86.7 & 100.0 & 68.2 & 95.5 & 81.8 \\
      \midrule
      \textbf{\textit{Overall}} & 100.0 & 83.3 & 95.7 & 80.9 & 88.4 & 83.7 \\
      \bottomrule
    \end{tabular}
\end{table}

%% file: tables/results/RQ3/rq3_shape_rsr.tex
\begin{table}[h]
  \centering
  \caption{\textit{Shape-Mismatch Error Repair Success Rate} (\textit{ShapeRSR},\%) of \textsc{DeepGroundRepair} and Variant~3 on programs with initial tensor-shape mismatch failures. \textsc{DeepGroundRepair} and Variant~3 are denoted by \textsc{DeepGR} and V3, respectively. ``Change'' denotes the performance difference between \textsc{V3} and \textsc{\textsc{DeepGR}}. The ``\textcolor{red}{$\downarrow$}'' and ``\textcolor{blue}{$\uparrow$}'' indicate lower and higher \textit{ShapeRSR}, respectively.}

  \label{tab:rq3_shape_rsr}
  \resizebox{\linewidth}{!}{
  \begin{tabular}{
    @{\extracolsep{\fill}}
    ll
    ccc
    @{\hspace{8pt}}
    ccc
    @{\hspace{8pt}}
    ccc
    @{}
  }
    \toprule
    & & \multicolumn{3}{c}{\textbf{TensorFlow}}
    & \multicolumn{3}{c}{\textbf{PyTorch}}
    & \multicolumn{3}{c}{\textbf{PaddlePaddle}} \\
    \cmidrule(lr){3-5}
    \cmidrule(lr){6-8}
    \cmidrule(l){9-11}
    \multirow[c]{-2.2}{*}{\textbf{Model}}
    & \multirow[c]{-2.2}{*}{\textbf{Error Type}}
    & \textsc{V3} & \textsc{DeepGR} & Change
    & \textsc{V3} & \textsc{DeepGR} & Change
    & \textsc{V3} & \textsc{DeepGR} & Change \\
    \midrule
    \multirow{2}{*}{Gemma-12B}
    & Intra-path
    & 16.7 & 0.0  & \textcolor{red}{$\downarrow$16.7}
    & 18.2 & 56.8 & \textcolor{blue}{\textbf{$\uparrow$38.6}}
    & 14.3 & 38.1 & \textcolor{blue}{\textbf{$\uparrow$23.8}} \\
    & Inter-path
    & 6.2  & 56.2 & \textcolor{blue}{\textbf{$\uparrow$50.0}}
    & 0.0  & 0.0  & --
    & 0.0  & 0.0  & -- \\
    \midrule
    \multirow{2}{*}{Mistral-14B}
    & Intra-path
    & 0.0  & 25.0 & \textcolor{blue}{\textbf{$\uparrow$25.0}}
    & 14.9 & 53.2 & \textcolor{blue}{\textbf{$\uparrow$38.3}}
    & 9.3  & 51.2 & \textcolor{blue}{\textbf{$\uparrow$41.9}} \\
    & Inter-path
    & 7.9  & 31.6 & \textcolor{blue}{\textbf{$\uparrow$23.7}}
    & 0.0  & 0.0  & --
    & 0.0  & 0.0  & -- \\
    \bottomrule
  \end{tabular}
  }
\end{table}

%% file: figures/plot/RQ3/sankey.tex
\begin{figure}[h]
    \centering
    \includegraphics[width=\linewidth]{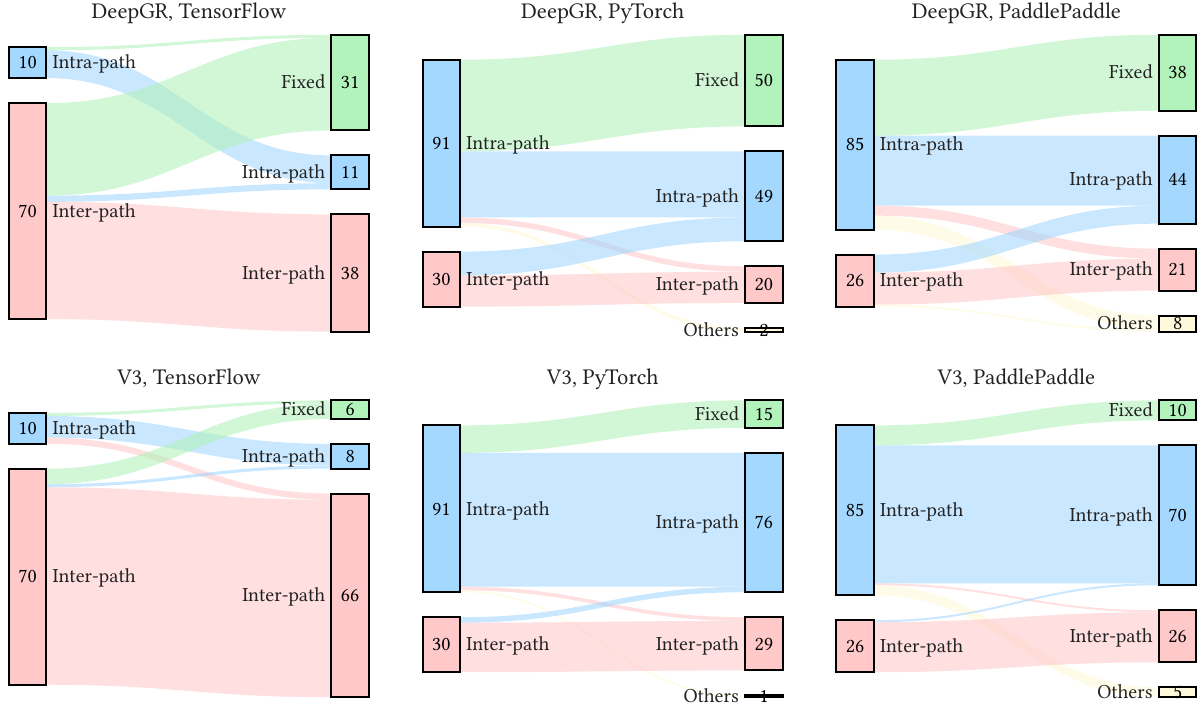}
    \caption{Transition distributions of tensor shape mismatch failures from initial categories to final outcomes, aggregated across Gemma-12B and Mistral-14B. In the figure, \textsc{DeepGroundRepair} and Variant~3 are denoted by \textsc{DeepGR} and V3, respectively. ``Fixed'' indicates successful repair, while ``Others'' captures a small number of corner cases beyond the intra-path/inter-path taxonomy.}
    \label{fig:root_cause_transition_sankey}
\end{figure}

%% file: figures/plot/RQ4/RQ4.tex
\begin{figure}[h]
    \centering
    \includegraphics[width=\linewidth]{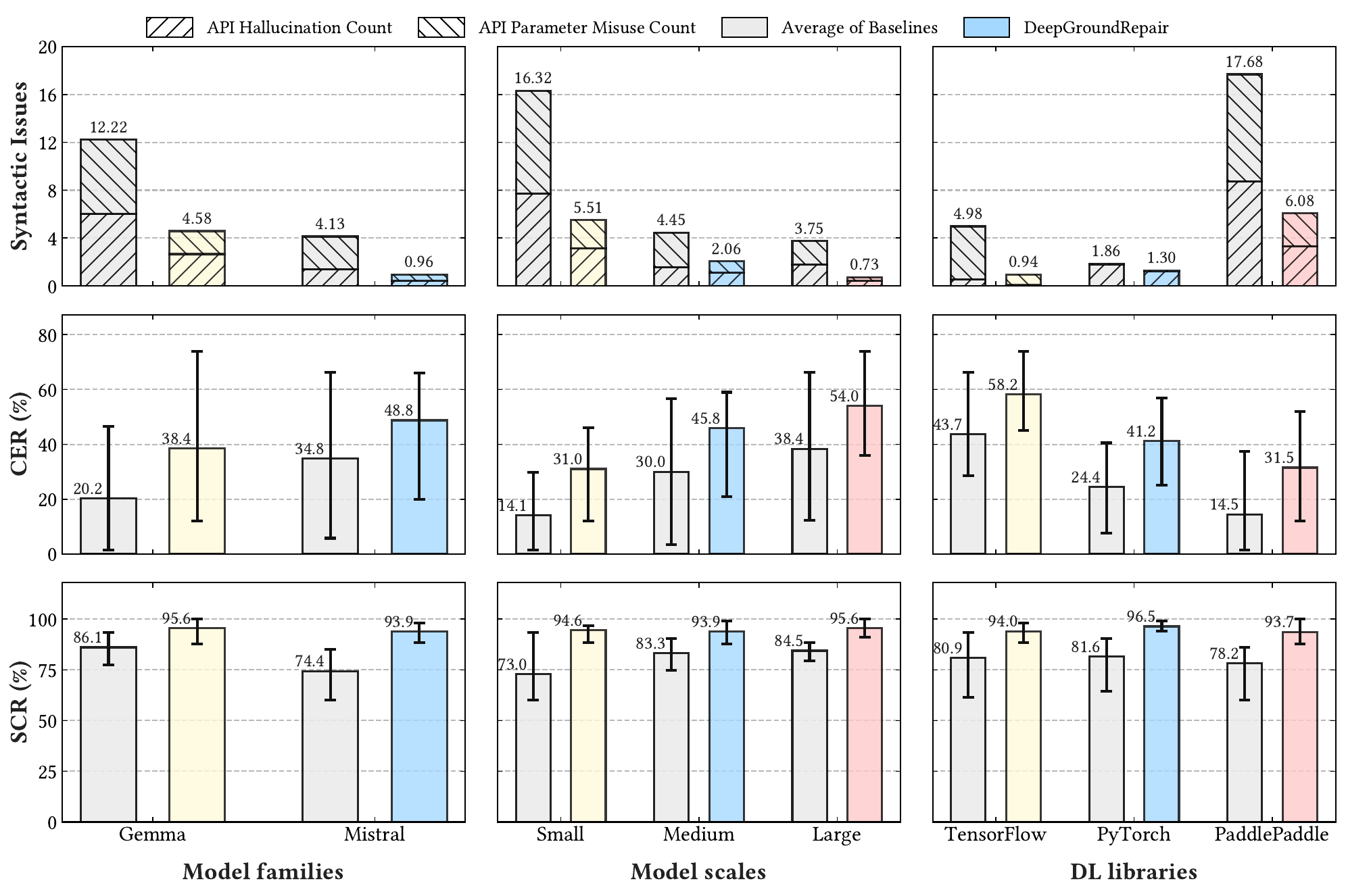}
    \caption{Generalizability of \textsc{DeepGroundRepair} across model families, model scales, and DL libraries. Each bar pair compares the group-level averages of \textsc{DeepGroundRepair} with the corresponding baseline mean. Hatching in the syntactic-issue row distinguishes API hallucination counts from API parameter misuse counts. Whiskers in the \textit{Code Executability Rate} (\textit{CER}) and \textit{Semantic Consistency Rate} (\textit{SCR}) rows show the minimum and maximum configuration-level values within each group.}
    \label{fig:generalizability_matrix}
\end{figure}

%% file: tables/results/RQ5/RQ5_tokens.tex

\begin{table}[h]
  \centering
  \caption{Average \textit{Token Cost} (in thousands of tokens) by model, DL library, and method. Here, \textsc{DeepGR} denotes \textsc{DeepGroundRepair}, and B1--B4 denote Baselines~1--4, respectively. ``$\Delta$ (\%)'' reports the relative change of \textsc{DeepGR} over each corresponding baseline using the values in the corresponding ``Overall Avg.'' row.}
  \label{tab:rq5_token_cost_b3b4_conditional}
  \setlength{\tabcolsep}{3pt}
  \resizebox{\linewidth}{!}{%
    \begin{tabular}{
    @{\extracolsep{\fill}}
    @{}
    l
    ccccc
    @{\hspace{5pt}}
    ccccc
    @{\hspace{5pt}}
    ccccc
    @{}
  }
    \toprule
    &
    \multicolumn{5}{c}{\textbf{TensorFlow}}
    & \multicolumn{5}{c}{\textbf{PyTorch}}
    & \multicolumn{5}{c}{\textbf{PaddlePaddle}} \\
    \cmidrule(lr){2-6}
    \cmidrule(lr){7-11}
    \cmidrule(lr){12-16}

    \multirow[c]{-2.2}{*}{\textbf{Model}}
    & B1 & B2 & B3 & B4 & \textsc{DeepGR}
    & B1 & B2 & B3 & B4 & \textsc{DeepGR}
    & B1 & B2 & B3 & B4 & \textsc{DeepGR} \\
    \midrule

    Gemma-4B
    & 0.53 & 0.54 & 1.09 & 1.10 & 1.10
    & 0.61 & 0.61 & 1.23 & 1.23 & \textbf{0.97}
    & 0.72 & 0.57 & 1.44 & 1.12 & \textbf{1.12} \\

    Gemma-12B
    & 0.75 & 0.70 & 1.56 & 1.43 & \textbf{1.28}
    & 0.76 & 0.73 & 1.55 & 1.50 & \textbf{1.15}
    & 0.76 & 0.70 & 1.54 & 1.41 & \textbf{1.07} \\

    Gemma-27B
    & 0.65 & 0.65 & 1.36 & 1.34 & \textbf{1.17}
    & 0.76 & 0.75 & 1.56 & 1.54 & \textbf{1.11}
    & 0.77 & 0.74 & 1.56 & 1.48 & \textbf{1.11} \\
    
    \rowcolor{cyan!12}
    \textit{Average}
    & 0.64 & 0.63 & 1.34 & 1.29 & \textbf{1.18}
    & 0.71 & 0.70 & 1.45 & 1.42 & \textbf{1.08}
    & 0.75 & 0.67 & 1.51 & 1.34 & \textbf{1.10} \\

    \midrule
    \addlinespace[5pt]

    Mistral-3B
    & 1.35 & 0.91 & 2.09 & 1.80 & \textbf{1.50}
    & 1.42 & 0.96 & 2.18 & 2.06 & \textbf{1.33}
    & 1.51 & 0.98 & 2.21 & 2.04 & \textbf{1.58} \\

    Mistral-14B
    & 0.92 & 0.95 & 1.93 & 1.74 & \textbf{1.17}
    & 1.02 & 1.04 & 1.98 & 2.20 & \textbf{1.35}
    & 1.01 & 1.02 & 2.11 & 2.12 & \textbf{1.17} \\

    Mistral-24B
    & 0.80 & 0.62 & 1.54 & 1.35 & \textbf{0.89}
    & 0.94 & 0.66 & 1.73 & 1.44 & \textbf{0.96}
    & 0.75 & 0.75 & 1.56 & 1.55 & \textbf{0.85} \\
    
    \rowcolor{cyan!12}
    \textit{Average}
    & 1.02 & 0.83 & 1.85 & 1.63 & \textbf{1.19}
    & 1.13 & 0.89 & 1.96 & 1.90 & \textbf{1.21}
    & 1.09 & 0.92 & 1.96 & 1.90 & \textbf{1.20} \\

    \midrule
    \addlinespace[5pt]
    
    \rowcolor{cyan!12} 
    \textbf{\textit{Overall Avg.}}
    & 0.83 & 0.73 & 1.60 & 1.46 & \textbf{1.19}
    & 0.92 & 0.79 & 1.71 & 1.66 & \textbf{1.15}
    & 0.92 & 0.79 & 1.74 & 1.62 & \textbf{1.15} \\
    
    \rowcolor{cyan!12}
    \textbf{\textit{$\Delta$ (\%)}}
    & \textcolor{red}{$\uparrow$\textbf{43.4}}
    & \textcolor{red}{$\uparrow$\textbf{63.0}}
    & \textcolor{blue}{$\downarrow$\textbf{25.6}}
    & \textcolor{blue}{$\downarrow$\textbf{18.5}}
    & --
    & \textcolor{red}{$\uparrow$\textbf{25.0}}
    & \textcolor{red}{$\uparrow$\textbf{45.6}}
    & \textcolor{blue}{$\downarrow$\textbf{32.7}}
    & \textcolor{blue}{$\downarrow$\textbf{30.7}}
    & --
    & \textcolor{red}{$\uparrow$\textbf{25.0}}
    & \textcolor{red}{$\uparrow$\textbf{31.6}}
    & \textcolor{blue}{$\downarrow$\textbf{33.9}}
    & \textcolor{blue}{$\downarrow$\textbf{29.0}}
    & -- \\

    \bottomrule
  \end{tabular}
  }
\end{table}

%% file: figures/plot/RQ5/RQ5.tex
\begin{figure}[h]
    \centering
    \includegraphics[width=\linewidth]{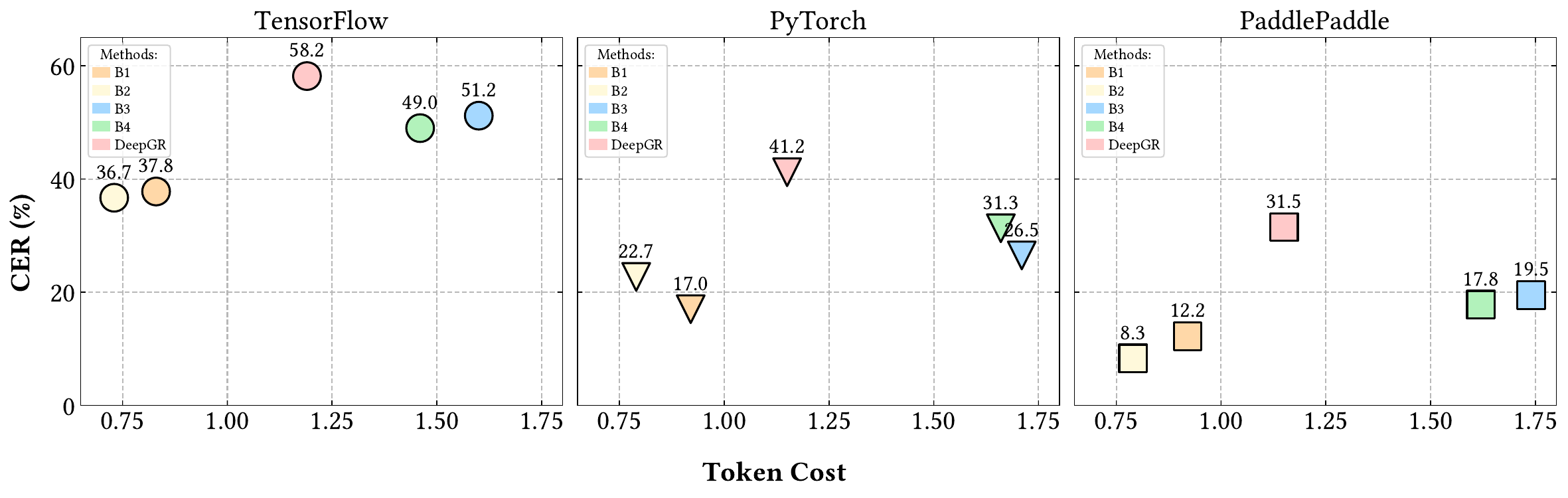}
    \caption{Scatter plots comparing \textit{Token Cost} (in thousands of tokens) and \textit{Code Executability Rate} (\textit{CER}, \%) across methods and DL libraries. Each point represents the average \textit{Token Cost} and \textit{CER} of a method across the six evaluated models for a given DL library. \textsc{DeepGR} denotes \textsc{DeepGroundRepair}, while B1--B4 denote Baselines~1--4, respectively.}
    \label{fig:token_cost_vs_cer}
\end{figure}

%% file: prompts/stage1_1.tex
\begin{figure}[h]
  \centering
  \includegraphics[width=0.68\linewidth]{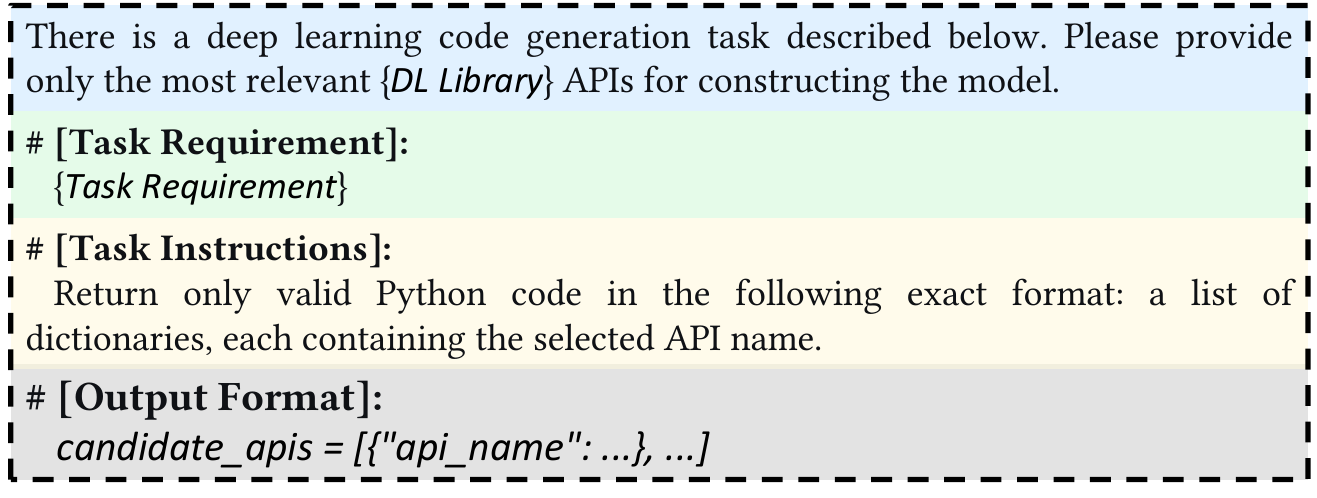}
  \caption{Prompting template of API recommendation in stage 1.}
  \label{fig:prompt1}
\end{figure}

%% file: prompts/stage1_2.tex
\begin{figure}[h]
  \centering
  \includegraphics[width=0.68\linewidth]{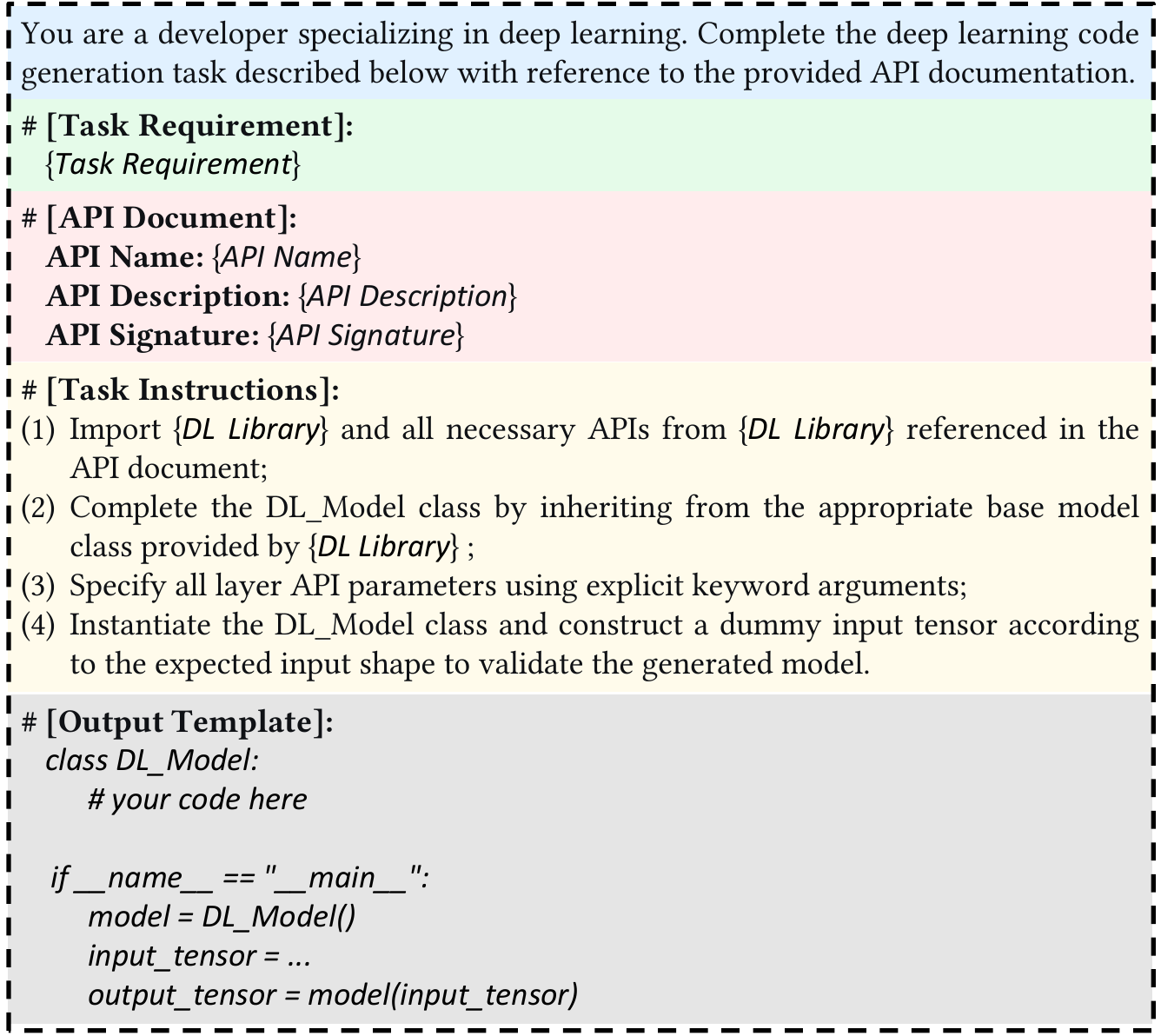}
  \caption{Prompting template of draft code generation in stage 1.}
  \label{fig:prompt2}
\end{figure}

%% file: prompts/stage3_1.tex
\begin{figure}[H]
  \centering
  \includegraphics[width=0.78\linewidth]{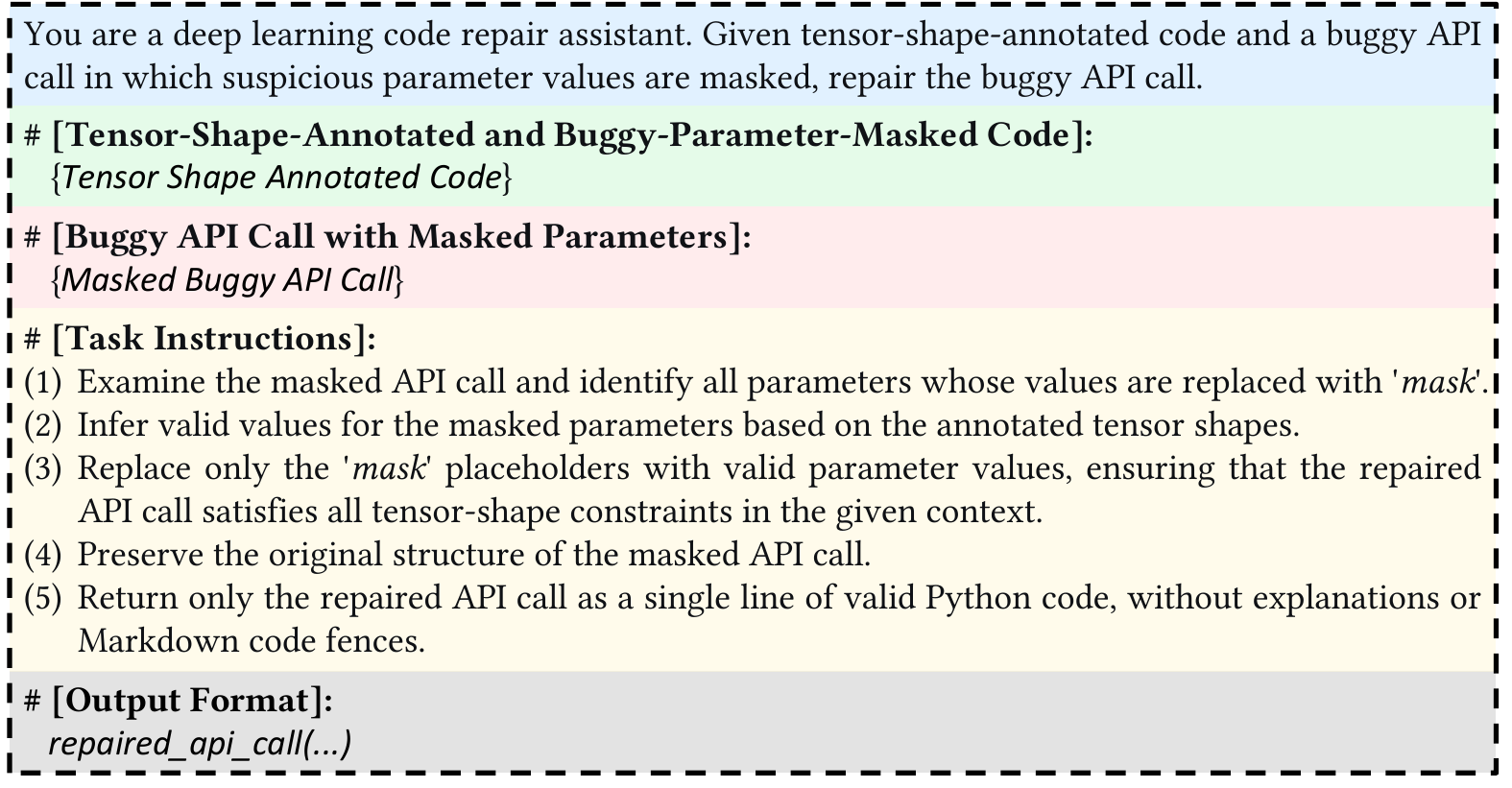}
  \caption{Prompting template of intra-path mismatch repair in stage 3.}
  \label{fig:prompt3}
\end{figure}

%% file: prompts/stage3_2.tex
\begin{figure}[H]
    \centering
    \includegraphics[width=0.78\linewidth]{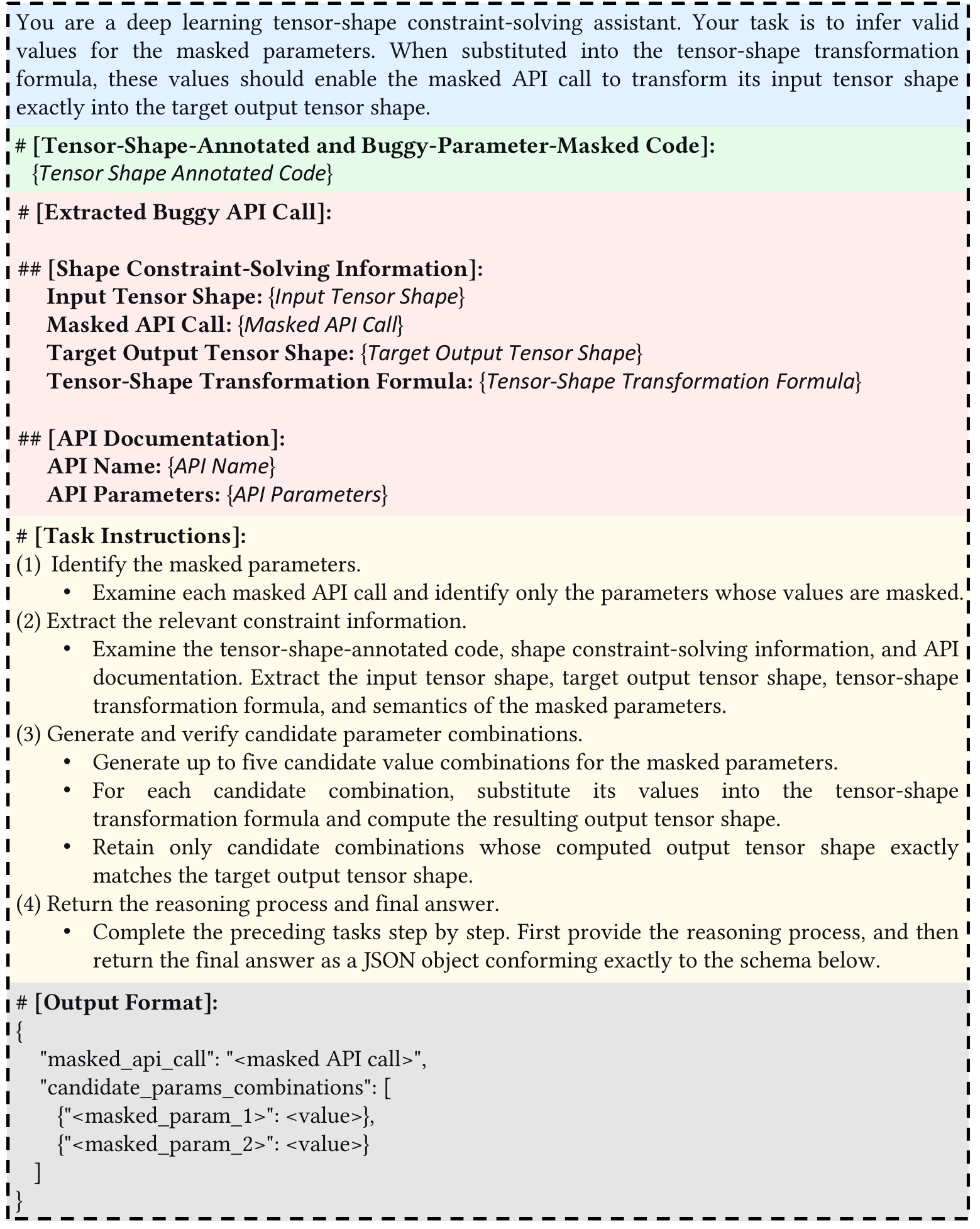}
    \caption{Prompting template of inter-path mismatch repair in stage 3.}
    \label{fig:prompt4}
\end{figure}